\documentclass[preprint2,trackchanges,lineno]{aastex631}

\hypersetup{linkcolor=red,citecolor=green,filecolor=cyan,urlcolor=magenta}

\usepackage{amsmath}

\renewcommand{\edit}[2]{#2}

\newcommand{\He}[1]{\ensuremath{{}^{#1}{\rm He}}} 
\newcommand{\C}[1]{\ensuremath{{}^{#1}{\rm C}}} 
\newcommand{\N}[1]{\ensuremath{{}^{#1}{\rm N}}}
\newcommand{\Ox}[1]{\ensuremath{{}^{#1}{\rm O}}}
\newcommand{\Ne}[1]{\ensuremath{{}^{#1}{\rm Ne}}}
\newcommand{\Na}[1]{\ensuremath{{}^{#1}{\rm Na}}}

\newcommand{\Ni}[1]{\ensuremath{{}^{#1}{\rm Ni}}}

\newcommand{\Si}[1]{\ensuremath{{}^{#1}{\rm Si}}}
\newcommand{\Fe}[1]{\ensuremath{{}^{#1}{\rm Fe}}}

\newcommand\Msun{\ensuremath{M_\sun}}

\accepted{September 1, 2023}

\shorttitle{Deflagrations of Hybrid White Dwarfs}
\shortauthors{Feldman et al.}
\graphicspath{{./}{figures/}}

\begin{document}
\nolinenumbers
\title{Dimming the Lights: 2D Simulations of Deflagrations of Hybrid C/O/Ne White Dwarfs using FLASH}

\correspondingauthor{Catherine Feldman}
\email{catherine.feldman@stonybrook.edu}

\author[0000-0001-7003-4431]{Catherine Feldman}
\affiliation{Department of Physics and Astronomy \\
Stony Brook University \\
Stony Brook, NY 11794-3800, USA}
\affiliation{Institute for Advanced Computational Science \\
Stony Brook University \\
Stony Brook, NY 11794-3800, USA}

\author[0000-0003-4483-7171]{Nathanael Gutierrez}
\affiliation{School of Computer Science \\
Georgia Institute of Technology \\
Atlanta, GA 30332, USA}

\author[0009-0007-0974-1891]{Ellis Eisenberg}
\affiliation{Yale University \\
New Haven, CT 06520, USA}

\author[0000-0003-2300-5165]{Donald E. Willcox}
\affiliation{Center for Computational Sciences and Engineering \\
Lawrence Berkeley National Lab \\
Berkeley, CA 94720, USA}

\author[0000-0002-9538-5948]{Dean M. Townsley}
\affiliation{Department of Physics and Astronomy \\
The University of Alabama \\
Tuscaloosa, AL 35487-0324, USA}

\author[0000-0001-5525-089X]{Alan C. Calder}
\affiliation{Department of Physics and Astronomy \\
Stony Brook University \\
Stony Brook, NY 11794-3800, USA}
\affiliation{Institute for Advanced Computational Science \\
Stony Brook University \\
Stony Brook, NY 11794-3800, USA}



\begin{abstract}
The dimmest and most numerous outlier of the Type Ia supernova population, Type Iax events, is increasingly being found in the results of observational campaigns. There is currently no single accepted model to describe these events. This 2D study explores the viability of modeling Type Iax events as a hybrid C/O/Ne white dwarf progenitor undergoing a deflagration using the multi-physics software FLASH. This hybrid was created using the stellar evolution code MESA, and its C-depleted core and mixed structure have demonstrated lower yields than traditional C/O progenitors in previous deflagration-to-detonation studies. To generate a sample, 30 ``realizations'' of this simulation were performed, the only difference being the shape of the initial matchhead used to start the deflagration. As consistent with earlier work, these realizations produce the familiar hot dense bound remnant surrounded by sparse ejecta. Our results indicate the majority of the star remains unburned ($\sim$ 70\%) and bound ($>$ 90\%). Our realizations produce total ejecta yields on the order of $10^{-2}$ - $10^{-1}$ \Msun, ejected \Ni{56} yields on the order of $10^{-4}$ - $10^{-2}$ \Msun, and ejecta kinetic energies on the order of $10^{48}$ - $10^{49}$ ergs. Compared to yields inferred from recent observations of the dimmest Type Iax events - SN 2007qd, SN 2008ha, SN 2010ae, SN 2019gsc, SN 2019muj, SN 2020kyg, and SN 2021fcg - our simulation produces comparable \Ni{56} yields, but too-small total yields and kinetic energies. Reignition of the remnant is also seen in some realizations.
\end{abstract}




\section{Introduction} \label{sec:intro}
Type Ia supernovae (SNe Ia) are ``standardizeable'' candles \textemdash their lightcurves align after applying the Phillips relation \citep{Phillips_1993}, which relates their maximum brightness and decline rate. This hallmark makes SNe Ia a tape measure for the universe, a property that has been used to demonstrate that the universe is expanding at an accelerating rate \citep{Riess_1998, Perlmutter_1999} and to provide a more precise calculation of cosmological parameters to test dark energy theories \citep{Abbott_2019}, among others. The accuracy of these distance measurements, and therefore the accuracy of their applications, depends on how well understood the explosion mechanism is; not all SNe Ia are created equal.

SNe Ia are thermonuclear explosions of a white dwarf (WD), a burned-out star that is subject to a mass limit, the Chandrasekhar mass ($M_\text{Ch}$). Upon obtaining extra mass from a companion star, the WD reignites fusion reactions and subsequently explodes if it is composed principally of C/O. The mass limit of a WD greatly limits the amount of material that can be ejected from the system, including the radioactive \Ni{56} whose decay gives the event its brightness.

Although the resulting light and elemental composition of a supernova are able to be observed, both the progenitor model and the explosion mechanism remain elusive. These are the keys to understanding their diversity, which indicates that the initial WD (progenitor) does not always have the same mass or elemental composition, and may have different processes causing its explosion (explosion mechanism). Especially interesting are the outliers of this population, which do not follow the Phillips relation. The dimmest and most numerous of these outliers are categorized into their own class: Type Iax events.

This paper is organized as follows. Section \ref{sec:intro} describes the background and characteristics of SNe Ia and their dim subclass Type Iax (Section \ref{subsec:iax}), as well as the proposed explosion mechanism (Section \ref{subsec:deflagration}) and progenitor system (Section \ref{subsec:progenitors}) used for this study. A brief review of related previous work is given in Section \ref{subsec:prevStudies}. We then describe our methodology in detail in Section \ref{sec:methods}, including the creation of our progenitor model (Section \ref{subsec:WD}); simulation of the explosion (Section \ref{subsec:FLASH} and Appendix \ref{ap:sim}); and formation of a sample (Section \ref{subsec:suite}). Section \ref{sec:analysis} shows how we calculate the yields of our simulations. The results are shown in Section \ref{sec:results} and their implications are discussed in Section \ref{sec:conclusion}. Finally, Section \ref{sec:future} describes the next steps we will take to perform our future 3D study.

\subsection{Type Iax Events} \label{subsec:iax}
A subset of SNe Ia were first classified under the name Type Iax by \citet{Foley_2013}. Also called “02cx-like" events as a homage to their characteristic member, this group encompasses 5 - 31\% of all SNe Ia, and contains over 50 members, of which the most recently analyzed is SN 2021fcg \citep{SN2021fcg_Karambelkar}. Type Iax events can be distinguished from standard SNe Ia by their lower and wider range of peak magnitudes (-19 $< M_{V, \text{peak}} <$ -13), reduced ejecta velocities (2000 $< v_{\rm{ej}} <$ 8000 km/s) and explosion energies ($\sim 10^{49}-10^{51}$ ergs), slower decline rates in the redder band lightcurves, the absence of a second maximum in the near infrared band lightcurve, significant ejecta mixing, slow spectral evolution, and divergence from the Phillips relation \citep{Jha_2017}. Type Iax events also produce much smaller ejecta yields, inferred from observations to be on average 0.5 \Msun, $\sim$ 0.003 – 0.3 \Msun ~of which is \Ni{56} \citep{Foley_2013} (see Section \ref{subsec:obs} for observational yields). 
 As these characteristics make Type Iax events substantially different than SNe Ia, it has been suggested that their explosion mechanisms are different. As described in the next section, some leading models for a SN Ia involve a combination of a deflagration (subsonic) and a detonation (supersonic), while for Type Iax events, the hypothesized explosion mechanism is a deflagration alone.

\subsection{Explosion Mechanism: Deflagration} \label{subsec:deflagration}

Deflagrations have been suggested as an explosion mechanism that could reproduce the dimness and small ejected masses of Type Iax (See Section \ref{subsec:iax}) events \citep{Foley_2013, Jha_2017}. A deflagration evolves by conduction-induced burning and is therefore subsonic, meaning that the flame propagates slowly enough that the star has time to react, typically by expansion. This expansion reduces the fuel density, which causes the burning to extinguish before consuming the entire star. The result is a slower, less violent process than either a detonation alone or a deflagration-to-detonation transition (DDT), both of which involve supersonic burning \edit1{that} can consume the star in a few tenths of a second. The gentler deflagration mechanism is expected to produce much less \edit1{iron-group element (IGE) dominated material}, and therefore much less \Ni{56}, than a detonation or a DDT, and will accordingly produce a dimmer event. In similar studies, this deflagration explosion mechanism has been referred to as a `failed' \citep{Foley_2013} or `pure' \edit1{\citep{Leung_2020,Lach_2021}} deflagration.

One approach to creating a deflagration in a WD is the single-degenerate picture, in which a WD accretes mass from a companion star until thermonuclear runaway occurs. The composition and mass of the companion as well as the accretion method are variable, but the combination must produce an accretion rate that allows the WD to grow in mass, either by creating conditions for stable burning or by means of multiple nova explosions that leave some of the accreted mass behind. Interestingly, the one and only progenitor system that has ever been observed, completely serendipitously, for a Type Iax event is that of a He-donor companion star for SN 2012Z, giving credence to this hypothesis \citep{McCully_2014, McCully_2022}.

The single-degenerate picture has not been able to explain the full variation in Type Iax events. An alternative being explored is the double-degenerate picture, where a deflagration is ignited by the accretion disk of a WD-WD merger \citep{SN2021fcg_Karambelkar}. There is also observational evidence in support of the double-degenerate picture: the only observed Type Iax event remnant, the infrared source 2MASS J00531123+6730023 or ``Parker's star'', has been argued to be the result of a C/O + O/Ne WD-WD merger \citep{Gvaramadze_2019, Oskinova_2020, Ritter_2021}. This study's focus is placed on the single-degenerate picture.

The end state of simulations of deflagrations often shows a two-part system; as the deflagration is not strong enough to unbind the entire star, a large part of the star remains bound and is mixed with burning ashes, while the rest of the mass is energetic enough to be ejected from the system \edit1{\citep{Kromer_2015, Leung_2015, Lach_2021}}. Section \ref{subsec:prevStudies} gives an overview of the most recent of these studies in the single-degenerate, centrally ignited, near-Chandrasekhar mass scenario. Simulations of traditional C/O WD progenitors undergoing deflagrations produce an amount of \Ni{56} consistent with high and intermediate brightness Type Iax events. One suggested method to reach the dimmest events is to instead use a hybrid C/O/Ne WD progenitor.

\subsection{Hybrid Progenitors} \label{subsec:progenitors}
White dwarfs are inactive cores --- the end state of low mass (\edit1{$\lesssim$ 8 - 10 \Msun)} stars. Their lighter outer layers have either dissipated as a \edit1{planetary} nebula or, if the star is in a binary pair, been pulled away by their companion. What is left behind is a dense core supported by electron degeneracy pressure –- this is a white dwarf (WD). The WD’s elemental composition is governed by the mass of the initial star, which determines how far down the chain fusion can occur. Stars with masses $<$ 8 \Msun~ form carbon-oxygen (C/O) WDs, while heavier ones (8 - 10.5 \Msun) create oxygen-neon (O/Ne) WDs. These C/O WDs will produce a SNe Ia under the right accretion conditions, while O/Ne WDs will collapse before ignition can occur. \citep{Hansen_2004}.

Recent research has postulated a new ``hybrid'' C/O/Ne model \edit1{ -- a WD made of primarily \C{12}, \Ox{16}, and \Ne{20} whose interior has homogenized due to mixing}. These hybrids are thought to form when the WD's carbon burning is quenched by convective boundary mixing before it reaches the core \edit1{, creating} a C-rich C/O core surrounded by a O/Ne shell \citep{Denissenkov_2013}.
\edit1{It has been demonstrated that these layers mix as the WD cools due to the electron-to-baryon ratio ($Y_e$) gradient between the heavier O/Ne/Na mantle and the lighter C/O core \citep{Brooks_2017}. The burned material has a smaller $Y_e$ due to the beta decay of \N{13}
that occurs during the burning. This gradient becomes unstable to convection as the WD cools: as heat is taken away from the mantle through neutrino transport and conduction, the mantle starts to sink and mix with the core material. This convective mixing reduces the $Y_e$ gradient, which moves to a \edit1{uniform} value as the system becomes isothermal. Thermohaline convective mixing continues the process until the entire \edit1{interior of the} star is fully mixed and the $Y_e$ gradient is 0, producing a flat composition profile.}

\edit1{The relatively short cooling times allow for the WD's interior to be fully mixed before the onset of accretion. This mixing significantly decreases the carbon fraction in the core. This homogenization due to mixing has been demonstrated in both 1D stellar evolution calculations by \citet{Brooks_2017} and 2D direct numerical simulation by \citet{Schwab_2019}.}

\edit1{
In this paper, we refer to a ``hybrid WD'' as a C/O/Ne WD that has completely mixed during cooling.
}
The hybrid progenitor used for this study \edit1{is described in detail in Section \ref{subsec:WD} } and has been shown to produce lower \Ni{56} and ejecta yields in DDT simulations due to its lower carbon fraction \citep{Willcox_2016, Augustine_2019}.

\subsection{Previous Deflagration Studies} \label{subsec:prevStudies}
There are several studies that perform full hydrodynamic simulations of deflagrations of WDs in the centrally-ignited near-Chandrasekhar mass scenario. The availability of computational resources that satisfy the intense requirements of such simulations, as well as the complexity of developing software that reproduces the behavior of a thermonuclear flame on an unresolved scale, are two of the main roadblocks in achieving this goal. Over the past decade, great strides have been made \edit1{in} both areas, and it is now possible to perform systematic studies that investigate not only nucleosynthetic yields and observables, but also the effects of the initial flame ignition. LEAFS \citep{Reinecke_1999_levelset, Reinecke_1999_flame, Reinecke_2002} and FLASH \citep{Fryxell_2000, Calder_2002} are two software systems capable of performing such 3D simulations, and a third by \citet{Leung_2015} is currently functional up to 2D studies. 

Here, we outline results from the most recent studies of \edit1{ deflagrations} in the literature from \edit1{the three} codes: \citet{Lach_2021} (LEAFS, C/O progenitor), \citet{Kromer_2015} (LEAFS, \edit1{unmixed} C/O/Ne progenitor), \edit1{\citet{Leung_2020} (\citet{Leung_2015}, C/O and mixed C/O/Ne progenitors)}, and \citet{Long_2014} (FLASH, C/O progenitor). \citet{Lach_2021} is the newest study to date, and gives \edit1{an excellent} contemporary overview of the state of the field. We briefly describe the current state of research here.

\subsubsection{C/O WDs}
The more traditional C/O WD progenitor has been simulated undergoing a deflagration by multiple studies. Here, we present the most recent studies that use \edit1{each of the three codes.}

\citet{Long_2014} performed deflagrations of C/O WDs using an \edit1{earlier} version \citep{Calder_2007, Townsley_2007} of the software used for this study, FLASH (see Section \ref{subsec:FLASH}). Their 1.365 \Msun~ progenitor was made of 50/50 C/O, and ignited by placing varying numbers (63 - 3500) of 16 km burned spherical regions within a central sphere of 128 - 384 km radius. The end state of the simulation is recognized as having a dense core surrounded by less dense material, however unlike most other studies, all of this material is considered to be ejecta. It is likely, especially since they observe material sinking towards the core, that only some of this material is actually ejected, while the majority recombines to form a bound remnant. This is corroborated by the fact that the generated light curves do not match those of  observations, though the simulation with the fewest number of ignition points is the closest. The amount of \Ni{56} in the core vs.\ the surrounding material is related to the number of ignition points used; simulations with fewer ignition points show the majority of \Ni{56} in the less dense surrounding material, while those with more ignition points show the majority of \Ni{56} trapped in the core.

\citet{Lach_2021} perform 3D simulations of deflagrations of C/O progenitors using LEAFS. It utilizes a parabolic piecewise method (PPM) hydrodynamics solver, a FFT-based gravity solver that solves the full Poisson equation, a subgrid turbulence model, and \citet{TimmesAndArnett_1999}'s Helmholtz EOS with Coulomb corrections. The flame front is tracked using a level-set method, and nuclear burning is approximated by a reduced reaction network of 5 pseudospecies ($\alpha$, \C{12}, \Ox{16}, IME, and IGE), where the density of the cell determines the burned composition. The NSE state is dynamic and is adjusted due to electron capture. It uses two nested expanding grids to simultaneously track the remnant and the ejecta. Elemental abundances were calculated from the temperature and density history of Lagrangian tracer particles added to the simulation using the nuclear network code YANN, and light curves were generated using the radiative transfer (RT) code ARTIS.

\citet{Lach_2021} use a traditional equal parts C/O progenitor with a central density of 2.6 $\times$ 10$^9$ g/cm$^3$, central temperature of 6 $\times$ 10$^8$ K, and solar metallicity. They ignite it in a single spot, using 8 overlapping spheres, each with a 5 km radius, placed 60 km off-center. To explore the parameter space, \citet{Lach_2021} vary the central density, metallicity, and ignition offset, as well as add a few rigidly rotating models and one with a C-depleted core.

\citet{Lach_2021} find the familiar final state of a remnant surrounded by ejecta. The majority of the system, 1.09 - 1.38 \Msun, remains bound. \citet{Lach_2021} suggest their findings could account for Type Iax events of intermediate and high brightness: they report ejected masses between 0.014 and 0.301 \Msun and ejected \Ni{56} masses between 0.0058 - 0.092 \Msun. They find that due to the density gradient inside the star, the deflagration only propagates outward from the ignition point, creating an asymmetry that gives a kickback velocity to the remnant. Depending on the magnitude of this velocity, the remnant may be ejected from the binary system.

\edit1{\citet{Leung_2020} perform 2D studies of a C/O WD (and also C/O/Ne WDs, described in the following section) undergoing a pure deflagration using their own code \citet{Leung_2015}. This code uses a $5^{\rm{th}}$-order weighted essentially non-oscillatory (WENO) hydrodynamics solver, a level-set method to track the flame front coupled to a subgrid turbulence model, a three-stage burning model, a multipole gravity solver, and a Helmholtz EOS with Coulomb corrections. A 19-isotope reaction network is used to calculate the nuclear energy release. They simulate a quarter of the star, which has a cylindrical plus reflection symmetry.}

\edit1{\citet{Leung_2020} use a traditional 1.37 \Msun ~ C/O WD with a central density of 3 $\times$ 10$^9$ g/cm$^3$, a central temperature of 9 $\times$ $10^9$ K, and solar metallicity  which is added to the model as a \Ne{22} mass fraction of 0.025. It is centrally ignited with an initially burned region with three perturbations, which are called ``fingers''. They explore the effect of changing the central density of the WD (0.50 - 9.00 $\times$ 10$^9$ g/cm$^3$), the C/O ratio (0.3, 0.6, and 1.0), and the initial ignition type (central 3-finger and off-center single bubble). Their simulations eject a large amount of material (though the realization with the smallest central density ejected no mass); 0.92 - 1.36 \Msun~ is ejected , 0.23 - 0.34 \Msun ~ of which is \Ni{56}. The ejecta velocity is on the order of 5000 - 8000 km/s. These simulations could account for the brightest Type Iax members, including 2005hk and 2012Z.}


\subsection{C/O/Ne WDs}
There are not many studies that use these newly hypothesized hybrid progenitors, and the only 3D study available to date is that of \citet{Kromer_2015}, which uses an unmixed hybrid. \citet{Leung_2020}'s 2D study uses a mixed hybrid.

\citet{Kromer_2015} performed 3D simulations of deflagrations of a parameterized \edit1{unmixed} hybrid progenitor using an earlier version of LEAFS. They use a monopole gravity solver, and do not explicitly state the equation of state.

Their progenitor is a parameterized \edit1{unmixed} hybrid WD with a 0.2 \Msun ~50/50 C/O core surrounded by a 1.1 \Msun, 50/47/3 \Ox{16}/\Ne{20}/\C{12} shell. A 0.1 \Msun ~accreted layer of 50/50 C/O is placed on top of this. The central density is 2.9 $\times 10^9$ g/cm$^3$. The resulting star is ignited off-center with an asymmetric 5 kernel ignition, and burning was suppressed in the O/Ne region. 
We \edit1{emphasize} that this parameterized model neglects the mixing that occurs during both cooling and simmering phases, which homogenizes the composition of the WD, typically up to 1.0 \Msun, before flame ignition. Also the suppression of burning in the outer shell ought to be expected to produce results different from the natural mixed state.

Their simulation results in a 1.39 \Msun ~bound remnant with 0.014 \Msun ~of ejecta. The ejecta is comprised of mostly unburned material, with 6.4 $\times 10^{-3}$ \Msun ~of IGE, roughly half of which is \Ni{56}. It has a low ejecta energy of 1.8 $\times 10^{48}$ ergs of kinetic energy. This simulation gives a smaller \Ni{56}-to-mass ratio than expected by observations, and also too fast rise and decline of the light curves. However, its ejected \Ni{56} yields fall within range of that of the dimmest Type Iax events.

\edit1{Unlike that of \citet{Kromer_2015}, the hybrid C/O/Ne WD used in \cite{Leung_2020} is mixed. It has a homogeneous composition of 20\% \C{12}, 50\% \Ox{16}, and 30\% \Ne{20} by mass, and then adds in solar metallicity as 2.5\% \Ne{22}. It has the same central density (3 $\times$ 10$^9$ g/cm$^3$) as their C/O model. They investigate the effect of changing the central density of the WD (1.00 - 9.00 $\times$ 10$^9$ g/cm$^3$) and the initial ignition type (central 3-finger and off-center single bubble). They again find ejected yields similar to that of the brightest Type Iax events: ejecta masses in the range of 0.96 - 1.30 \Msun, 0.18 - 0.36 \Msun ~of which is \Ni{56}. These values are on average slightly smaller than their C/O WD counterparts.}

\subsection{This Study}  \label{subsec:thisStudy}
This study seeks to determine if hybrid C/O/Ne WDs that mix during cooling undergoing deflagration in the single-degenerate, centrally ignited, near-Chandrasekhar mass scenario can reproduce the estimated yields of the dimmest Type Iax events. The main differences from previous works are the use of stellar evolution calculations to create the progenitor (Section \ref{subsec:WD}) rather than a simpler parameterized model, and use of an ADR model flame in FLASH (Section \ref{subsec:FLASH}), which is significantly different from the level-set method implemented in LEAFS \edit1{and the software of \citet{Leung_2015}. We note that an ADR scheme in FLASH has been applied by several groups of researchers. The version we implement has had considerable development (see Section \ref{subsec:flame} and Appendix \ref{ap:flame}) since an earlier implementation was applied to deflagration models of C/O WDs by \citet{Long_2014}}. The results \edit1{presented here} are also directly compared to the DDT model from \citet{Augustine_2019}, which uses our same initial hybrid progenitor and software, as well as to observational data. \edit1{The effect of the initial ignition region on the final yields is also explored.}

\section{Methodology} \label{sec:methods}
There are two main pieces needed for this study: a model of a hybrid C/O/Ne WD progenitor and software to perform the deflagration. Section \ref{subsec:WD} describes how the progenitor is formed from stellar evolution calculations using the software instrument Modules for Experiments in Stellar Astrophysics (MESA), and subsequently mapped to a 2D grid for use in simulation. Section \ref{subsec:FLASH} describes the software system, FLASH, used to simulate the deflagration. This includes a description of the most vital pieces: the equation of state (EOS) (Section \ref{subsec:eos}); a module to model nuclear burning and the spread of the flame (Section \ref{subsec:flame}); the initial conditions for ignition (Section \ref{subsec:initflame}); the initial parameters for the ``fluff" material used in lieu of vacuum (Section \ref{subsec:initparams}); and the criteria for refinement of the grid (Section \ref{subsec:refinement}). In Section \ref{subsec:suite}, we describe how a sample of 30 simulations (each called a ``realization'') was chosen \edit1{to investigate the range of final yields, and how a subset of realizations was selected to explore the effect of the initial flame size on these yields}.

\subsection{Creating a Hybrid Progenitor} \label{subsec:WD}
We used the same hybrid C/O/Ne WD profile as \citet{Augustine_2019}. This progenitor was created using the 1D stellar evolution software instrument Modules for Experiments in Stellar Astrophysics (MESA), version 8118, \citep{Paxton_2011, Paxton_2013, Paxton_2015}, by the following method, which evolves a zero age main sequence (ZAMS) star through the WD phase, followed by a period of cooling and accretion. 
A ZAMS star with a metallicity of $Z$=0.02 was evolved using the set of controls from \citet{Farmer_2015} and the 29 species network ``\texttt{sagb\_NeNa\_MgAl.net}''. This network notably includes the \C{12}(\C{12},p)\Na{23} reaction of the carbon-burning process, which has been shown by \citet{DeGeronimo_2022} to have a strong impact on the \Ne{20} mass fraction of the end state WD. The ZAMS star evolves along the main sequence until it uses up the hydrogen in its core, transitions to the red giant branch (RGB), burns through its core helium, becomes a super-asymptotic giant branch (SAGB) star, and then ignites carbon burning off-center from the C/O core. This off-center ignition produces a carbon flame that is quenched by convective boundary mixing before it reaches the star's center \citep{Denissenkov_2013}, leaving a core of unburned C/O fuel surrounded by a mantle of burned O/Na/Ne. The resulting configuration is a 1.09 \Msun~ hybrid C/O/Ne WD, with a 0.4 \Msun~ C/O core with a C fraction of 0.345, enclosed by a 0.69 \Msun~ O/Ne/Na mantle.

There is a $Y_e$ gradient between mantle and core (burned and unburned material), which in this model is $\Delta Y_e$ = -0.00235 \citep{Brooks_2017}. The burned material has a smaller $Y_e$ due to the beta decay of \N{13} that occurs during the burning, which motivates the choice of the extended ``\texttt{sagb\_NeNa\_MgAl.net}'' network.
\edit1{ 
As described in Section \ref{subsec:progenitors}, it has been demonstrated that this gradient causes the mantle and core to mix as the WD cools, reducing the central C fraction and homogenizing the WD's \edit1{ interior far before accretion and subsequent ignition would occur  \citep{Brooks_2017, Schwab_2019}. }
} 
This mixing significantly decreases the carbon fraction in the core from 0.345 to 0.1419.

To model the accretion phase, the WD was placed in a binary system with a 1.4 \Msun~ He star with a 3 hour orbital period, following the prescription of \citet{Brooks_2016}. The He star was also evolved from a ZAMS star with a metallicity of $Z$=0.02, except it used the 8 species network ``\texttt{basic.net}''. The accreted layer therefore is devoid of the heavier elements, but this isn't expected to change the deflagration dynamics, which occur in the center of the star at the temperature peak.

As the WD gains mass through accretion from its He-rich companion, it compresses and increases in temperature. The mixed central part of the WD becomes hot enough to start ``simmering'': C-burning activates, but convection is able to carry excess heat throughout the material and prevent thermonuclear runaway. The simmering decreases the C-fraction a bit further, to $\sim$ 0.139. Beyond that, the underlying accreted layers begin burning He, enriching the outer layers of the star with C and O. Finally, the newly accreted material rests on the surface. Accretion continued until the convection in the core could not disperse the excess heat fast enough and thermonuclear runaway was imminent; in MESA this was considered to be the point where the central temperature was on the order of $10^{8.9}$ K. The resulting star just before the deflagration has a mass of 1.36 \Msun, with a central density of $2.2 \times 10^9$ g/cm$^3$ and a central temperature of $8.2 \times 10^8$ K.

The resulting progenitor WD was mapped to a cylindrically symmetric 2D grid to use as the input for a deflagration simulation in FLASH. The code that performed this transformation can be found on github (see Software section), and the transformation is described in \citet{Willcox_2016} and was performed for \citet{Augustine_2019}.

First, the profile was mapped to a 1D, 4 km grid (the maximum resolution of the simulation, as described in Section \ref{subsec:FLASH}). Next, the 29 species in the MESA profile were condensed into 4 in order to be used with our burning model in FLASH: \C{12}, \Ox{16}, \Ne{20}, and \Ne{22}. The prescription for doing this is to keep the \C{12} mass fraction ($X_{\C{12}}$) the same, and then use \Ne{22} as a proxy for the neutron-rich nuclides by calculating its mass fraction ($X_{\Ne{22}}$) using $Y_e$. The remaining mass fractions of \Ox{16} ($X_{\Ox{16}}$) and \Ne{20} ($X_{\Ne{20}}$)  were kept in the same ratio as in the MESA profile.
\begin{subequations}
\begin{equation}
    X_{\C{12}}^{\rm FLASH} =X_{\C{12}}^{\rm MESA}
\end{equation}
\begin{equation}
    X_{\Ne{22}}^{\rm FLASH} = \max(22 (\frac{1}{2}-Y_e), 0.0)
\end{equation}
\begin{equation}
    \nonumber R = X_{\Ne{20}}^{\rm MESA}/X_{\Ox{16}}^{\rm MESA}\\
\end{equation}
\begin{equation}
    X_{\Ox{16}}^{\rm FLASH}=\frac{1-X_{\C{12}}^{\rm FLASH}-X_{\Ne{22}}^{\rm FLASH}}{R + 1}
\end{equation}
\begin{equation}
    X_{\Ne{20}}^{\rm FLASH} = R X_{\Ox{16}}^{\rm FLASH}
\end{equation}
\end{subequations}

\begin{figure}[ht!]
\includegraphics[width=\columnwidth]{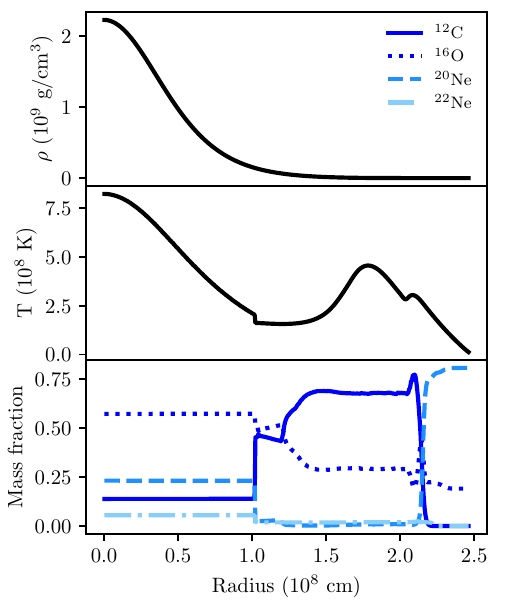}
\caption{Density (top), temperature (middle), and composition (bottom) of the end state of the hybrid C/O/Ne WD after accreting mass from its companion star. The 29 element reaction network was condensed into 4, so it can be used with our burning model in FLASH. Here, \Ne{22} is used as a proxy for neutron-rich material, and the accreted He has been added as \Ne{20} as they have the same electron-to-baryon ratio.}
\label{fig:initialWD}
\end{figure}

To force the grid to maintain hydrostatic equilibrium, the cell densities were adjusted. The density, temperature, and composition profile of the mapped model can be seen in Figure \ref{fig:initialWD}. 



Three ``layers'' can be seen, with transitions between them. The first is the composition-homogenized core convection zone (1.07 \Msun, 1.018$\times 10^8$ cm), which has enveloped the cooled, mixed star and further lowered its C-fraction. The expected adiabatic temperature profile can be seen in this layer. The next layer extends out to 2.15 $\times 10^8$ cm, and contains ashes from He-burning, where the accreted material has compressed and become hot enough (1.5-2.5 $\times 10^8$ K) to burn He. The remaining $10^{-6}$ \Msun ~layer that extends to the edge of the star (2.462 $\times 10^8$ cm) is made of unburned accreted material. Perhaps the most obvious difference between the MESA and FLASH models is the large amount of \Ne{20} in the accreted layer --- in reality, the accreted layer is mostly \He{4}, but this is not an element in the burning model. Since \Ne{20} and \He{4} have the same electron-to-baryon ratio, this should not change the equilibrium much. 



\subsection{Simulating a Deflagration} \label{subsec:FLASH}

The deflagration simulation was performed using \edit1{a customized version 
\citep{Calder_2007, Townsley_2007, Townsley_2009, Townsley_2016, Willcox_2016}} of FLASH, a modular, extensible, highly parallelizable software system \citep{Fryxell_2000, Calder_2002}. FLASH allows the combination of physics modules to create unique problems, which are then solved on a grid.

For this problem, we use FLASH version 4.6.2 with the PARAMESH adaptive mesh grid with a maximum resolution of 4 km; the split PPM hydro solver \citep{ColellaAndGlaz_1985}; the Poisson multipole gravity solver; a Helmholtz equation of state (EOS) with Coulomb corrections that describes the degenerate electron-positron plasma of the WD, overviewed in Section \ref{subsec:eos}; and \edit1{a burning model} specifically designed for modeling the thermonuclear processes in Type Ia supernovae that is described in Section \ref{subsec:flame}. For compilation, we used the GCC 10.2.0 compiler, MVAPICH version 2.2rc1 compiled with GCC, and parallel hdf5 version 1.8.20.

\subsubsection{Equation of State} \label{subsec:eos}
The EOS used for our application \edit1{describes the degenerate electron-positron plasma that makes up the WD.}
It is the Helmholtz EOS described in \citet{TimmesAndSwesty_2000} with additional Coulomb corrections. \edit1{Details of the implementation of this EOS are collected in Appendix \ref{ap:eos} and described in brief in \citet{Fryxell_2000}.}

As a deflagration produces slow-moving, sparse, and cool ejecta, it is important to be aware of the limits of this EOS. The Helmholtz EOS is meant to describe a fully ionized plasma, and its range of use has been cited as $10^{-10} < \rho < 10^{11}$ g/cm$^3$ and $10^4 < T < 10^{11}$ K \citep{TimmesAndSwesty_2000}. 
However, there is a region in $\rho$ -T space in which the Helmholtz EOS will return a negative internal energy. The bounds of this region are the right-triangle-shaped contour 
$E_{int} \sim 10^{11}$ ergs, which lies where 
$T < \sim10^6$ K and $\rho < \sim10^3$ g/cm$^3$. This region extends to lower densities as the temperature drops.
This problematic region is also present in the newest improvement of the Helmholtz EOS, the Skye EOS, which also describes a fully-ionized plasma over an even wider temperature and density range\citep{Skye_EOS}. \citet{Skye_EOS} find a triangular region of parameter space that fails when calculating the thermodynamic consistency relations, which they concisely explain: ``The feature at intermediate densities and low temperatures indicates negative pressures caused by our assumption of a fully ionized free energy in a region that should form bound states, indicating that Skye is not valid in that limit." \edit1{We discuss workarounds to this issue in Section \ref{subsec:initparams}.}

\subsubsection{\edit1{Burning} Model} \label{subsec:flame}
\edit1{The burning model has been developed from its original version in \citet{Calder_2007, Townsley_2007} to the current implementation of \citet{Townsley_2016}, and continues today.}
It is based upon the advection-diffusion-reaction (ADR) scheme of \citet{Khokhlov_1995} and \citet{Vladimirova_2006}. The addition of \Ne{20} to the model, which is necessary to simulate burning of this hybrid progenitor, is described in \citet{Willcox_2016}. Summarized below are the key characteristics, and more detail can be found in \edit1{Appendix \ref{ap:flame} and} the above references.

\edit1{A 3-stage burning process is used to model the evolution and energy release of a physical flame: carbon burning, burning to Si-group dominated nuclear statistical quasi-equilibrium (NSQE), and burning to Fe-group dominated nuclear statistical equilibrium (NSE). Each of these three burning stages is tracked by its own scalar progress variable -- $\phi_{fa}$ (fuel to ash), $\phi_{aq}$ (ash to NSQE), and $\phi_{qn}$ (NSQE to NSE), respectively -- that represents the fraction of a grid cell that has undergone that type of burning. It logically follows that $1 \geq \phi_{fa} \geq \phi_{aq} \geq \phi_{qn}$. $\phi_{fa}$ is evolved using an ADR scheme, and $\phi_{aq}$ and $\phi_{qn}$ at tuned timescales \citep{Calder_2007, Townsley_2016}. The properties of the final NSE state are evolved to account for changes in nuclear binding energy due to electron-capture as well as changes in composition due to hydrodynamic evolution.}

\edit1{\citet{Willcox_2016} found that the addition of \Ne{20} did not require any substantial changes to to the structure of this model. \Ne{20} is consumed in the first (C-burning) stage, and only serves to make the burning slower and less complete. The remainder of the stages progress with the same timescales as previously determined. Therefore, the only required change was to add \Ne{20} as an element in the composition of fuel for the first stage of the burning model.}

\edit1{Since the length scales of a realistic flame cannot be simulated inside a WD, the model flame is artificially thickened so that the flame front is exactly 4 grid zones wide. The minimum resolution needed for an accurate model is 4 km \citep{Townsley_2009}. The laminar flame speed is interpolated from a table of density, \C{12}-fraction, and \Ne{22} (a proxy for metallicity) fraction, and then boosted to account for unresolved Rayleigh-Taylor instability. To obtain laminar flame speeds for the lower densities and C-fractions of our hybrid progenitor, the table is simply extrapolated \citep{Willcox_2016}.}

\subsubsection{Ignition} \label{subsec:initflame}
As the actual flame cannot be resolved in this simulation, the ignition is approximated by a fully burned region whose surface is perturbed to capture the small-scale effects of convection. The deflagration was initialized at the temperature peak of the progenitor, at the center, with a 150 km radius “matchhead” of fully burned material ($\phi_{qn} = 1$). The shell of the matchhead was perturbed by an angular sinusoid function with a random harmonic number and amplitude, following the prescription of \citet{Townsley_2009}:
\begin{equation}
    r(\theta) = 150 \text{km} + 30 \text{km}\sum_{\ell=\ell_{min}}^{\ell_{max}}\sum_{m=-\ell}^{+\ell} A_\ell Y^m_\ell(\theta)
\end{equation}
The values of $\ell$ are taken from \citet{Townsley_2009}: $\ell_{max}$ was chosen so that that the perturbations are resolved, and a relatively large choice for $\ell_{min}$ allows the star time to expand and more accurately reproduce DDT observational yields. The solutions are limited to those that have only inward radial perturbations at the cylindrical symmetry axis, to prevent long thin plumes from forming and growing along the axis, a nonphysical behavior caused by our choice of cylindrical coordinates. In 2D simulations such as this, $m=0$ only is dictated by the assumed cylindrical symmetry. The sum reduces to:
\begin{equation} \label{eq:ignition}
    r(\theta) = 150 \text{km} + 30 \text{km}\sum_{\ell=12}^{16} A_\ell Y^0_\ell(\theta)
\end{equation}

These $A_\ell$ are set at the beginning of the simulation by a random seed. As will be described in Section \ref{subsec:suite}, for a statistical approach we run a suite of simulations, each of which has a different random seed that generates a unique set of $A_\ell$.

\subsubsection{Initial parameters} \label{subsec:initparams}
As the simulation is in 2D, a simplification that readily follows is to start with a zero background velocity field. In reality, the core of the star is convective, and future 3D simulations will include a velocity field derived from turbulence simulations \citep{Jackson_2014}, but as 2D simulations cannot describe turbulence, we neglect an initial velocity.

As the PPM hydrodynamics solver of FLASH requires material on every cell of the grid, the vacuum of space is instead replaced with a low-density ``fluff'' material surrounding the star \citep{Zingale_2002}. The fluff composition was set to match the abundances of the outermost zone of the initial profile, to eliminate any composition gradient. The fluff pressure was set to $5 \times 10^{14}$ dyne/cm$^2$, 25\% higher than the pressure at the edge of the star. 

The pressure of the outermost zone of the initial stellar profile is resolution-dependent. The steep edge of the surface of the star results in a resolution dependence of the surface pressure. We found that within the first timesteps of the simulation that the pressure at the edge of the star can jump up an order of magnitude from ${\sim} 4 \times 10^{14}$ dyne/cm$^2$ to ${\sim} 4 \times 10^{15}$ dyne/cm$^2$, which necessitated setting the fluff pressure to mitigate any resulting outwards pressure waves and subsequent inward shock due to this change. Since our deflagration is slow moving and the ejecta densities are sparse, we used a fluff density of $10^{-4}$ g/cm$^3$.

During the expansion of the ejecta after the explosion, some regions pass through the problematic region of our equation of state discussed in Section \ref{subsec:eos} near a density of $10^{-2}$ g/cm$^3$ and temperature of $10^6$ K. Interestingly, the areas where this negative internal energy are returned are the grid cells immediately in front of the plumes created by the flame, which have a much lower temperature (${\sim}10^6$ K) than the surrounding material. It is important to note that while this may be physical, it is likely due to an artifact of the piecewise parabolic split  hydrodynamics solver of FLASH. The cause of these low temperature regions will have to be further investigated.

In the meantime, to attempt to avoid this area of parameter space, we set the relatively high temperature floor of $2 \times 10^6$ K. This high temperature floor introduces a negligible amount of internal energy into our simulations at later times ($\sim$ 10 s). Since burning has stopped at this point and internal energy is not included in the bound vs unbound criteria, this increase in temperature floor does not significantly change the results. This is clearly an imperfect solution and some simulations were not able to successfully continue to the desired stopping time and were ended early.

\subsubsection{Refinement} \label{subsec:refinement}
We have chosen a fluff density on the order of $10^{-4}$ g/cm$^{3}$ and the ejecta reaches densities of the order $10^{-2}$ g/cm$^{3}$ or lower.
As a result, distinguishing between fluff and star material based on density, as was done in \citet{Townsley_2009}, becomes a less tenable strategy toward the end of our simulation. Instead, we replaced this with a combination of a fluff marker and a derefinement scheme. Adding a fluff marker was practically accomplished by initializing the simulation with a scalar variable, $0 \leq \phi_{\rm{fluff}} \leq 1$, where $\phi_{\rm{fluff}} = 0$ inside the star, and $\phi_{\rm{fluff}} = 1$ outside. This scalar is advected by the hydrodynamics much like a single species abundance, so that it effectively tracks the percentage of fluff in each grid cell. Rather than principally using density to detect non-star material, this marker was used to keep the fluff as derefined as possible, (neighboring blocks can only differ by 1 level of refinement at most), by setting the maximum refinement level of the fluff to 1 (4096 km), the largest grid cell size available for our simulation.

Burning stops at about 3 s of simulation time due to the decreasing density leading to the extinction of the propagating flame. The flame and dynamic ash were manually turned off at 5 s (some simulations burn faster or slower than others, so 5 s was chosen so as to not prematurely cut off the burning). The flame needed to be turned off because, as discussed in Section \ref{subsec:reignition}, the bound remnant in some simulations contracted to high enough densities that burning might occur. We have opted to consider such cases separately since the necessary conditions for reignition and the nature of the combustion under those conditions are uncertain. The densities and temperatures reached by the material in our simulation are low enough where continuing with dynamic ash enabled makes a negligible difference.

At the start of the simulation, the maximum refinement level was set to 11, which corresponds to a grid cell size of 4 km, the size necessary to accurately model the flame. The maximum refinement level of the non-energy generating regions (ie anything not burning) was set to 9 (16 km), and for fluff 1 (4096 km). After 3 s, for derefinement, we decreased the non-energy generating regions’ maximum refinement level by 1 (ie doubled the minimum grid cell size) every time the star doubled in size: 3 s (refinement level 8, 32 km), 5 s (refinement level 7, 64 km), and 10 s (refinement level 6, 128 km). Since after 3s burning has stopped, all regions are classified as non-energy-generating and are therefore subject to this refinement constraint.

\subsection{Suite} \label{subsec:suite}
To obtain results representative of a population, rather than a single member, we ran a suite of simulations with different initial flame perturbations, each of which we call a ``realization'' (see Section \ref{subsec:initflame}). It has been demonstrated by \citet{Calder_30realz} that 15-20 realizations of a DDT simulation are enough to cause the standard deviation of the \Ni{56} yield of the sample to converge\edit1{, however it is unclear whether this applies to deflagration simulations. Nevertheless, to follow the prescription of \citet{Augustine_2019} which used the same hybrid progenitor undergoing DDT}, thirty realizations (numbered R11- R40) were performed for this study. \edit1{We believe that this provides useful, though incomplete, sampling of possible outcomes.} 

To create a more direct comparison to WD undergoing DDT, each realization used the same random seed as its DDT counterpart in \citet{Augustine_2019}, so that the geometry of the initial flames were perfectly replicated. The initial conditions of the fluff and the refinement criteria, however, had to be adjusted from that of \citet{Augustine_2019} as described above in Section \ref{subsec:initparams} so that the deflagration could evolve for the requisite longer period of time. A description of how the data from these thirty realizations was analyzed can be found below in Section \ref{sec:analysis}, and a summary and comparison of the results of these thirty realizations is shown in the following Section \ref{sec:results}.

\edit1{To better understand the effect of the initial flame radius on the results, we performed additional simulations using 10 realizations from our sample that represented the full range of yields. The flame radius was changed from its original 150 km to 75 km (``half-radius'') and 300 km (``double-radius'').}

\section{Analysis} \label{sec:analysis}
Analysis was performed and plots were created using the yt package \citep{Turk_2010}. For each realization, \edit1{data was collected}
every 0.1 s of simulation time. To distinguish between the remnant and ejecta, we separate our results into regions of positive and negative \edit1{gravitational} binding energy, respectively, calculated by Equation \ref{eq:BE}. We choose this separation so that we can directly compare these simulated yields with those estimated from observational data, as discussed in Section \ref{subsec:obs}.

\begin{equation} \label{eq:BE}
    E_{\rm binding} = E_{\rm grav} + E_{\rm kinetic}
\end{equation}

\edit1{Here, $E_{\rm grav}$ is the gravitational potential energy, and $E_{\rm kinetic}$ is the kinetic energy (KE) of each grid cell.} Internal energy was not included in this calculation because the high temperature floor artificially increased the internal energy of the low density, low temperature regions at the edge of the plumes of burned material, which are often located near the line of 0 binding energy. The internal energy of these regions is orders of magnitude smaller than the kinetic or gravitational potential energy at late times, so the final yields of the remnant and ejecta are negligibly affected by excluding internal energy from the binding energy calculation.

Since our combustion model does not explicitly evolve abundances of particular nuclear species, as discussed in Section \ref{subsec:FLASH}, the yields of specific elements, namely IGEs and \Ni{56}, must be calculated using the progress variables ($\phi_{fa}$, $\phi_{aq}$, $\phi_{qn}$), neutron excess of the fluid ($Y_e$), and initial, unburned mass fractions of \C{12} ($X_{\C{12}}$), \Ne{20} ($X_{\Ne{20}}$), and \Ne{22} ($X_{\Ne{22}}$) the proxy for neutron-rich material).

The simulation considers the remaining material as $\Ox{16}$.
\begin{equation}
    X_{\Ox{16}} = 1 - X_{\C{12}} - X_{\Ne{20}} - X_{\Ne{22}}
\end{equation}

The mass of material in each grid cell was calculated using the mass fraction, density, and 3D volume of each cell.
\begin{equation}
    M_{i} = X_{i} \rho V
\end{equation}
where $M_i$ is the mass of substance $i$, $X_i$ is the mass fraction of substance $i$, $\rho$ is the density and $V$ the 3D annular volume of a cell.

The mass fraction of IGE is simply the progress variable $\phi_{qn}$, as described in \edit1{Section \ref{subsec:flame}}, since the difference between baryon density and mass density is unsubstantial.
\begin{equation}
    X_{IGE} = \phi_{qn}
\end{equation}

The calculation of the mass fraction of \Ni{56} present in each grid cell ($X_{\Ni{56}}$) follows the estimation of Equation 29 of \citet{Townsley_2009}, which assumes that \Fe{54} and \Ni{58} are the next most-common elements in the fully burned material.
\begin{equation} \label{eq:Ni}
    X_{\Ni{56}} = \max \left(\phi_{qn} \frac{Y_{e,n} - Y_{e,\Fe{54}+\Ni{58}}} {Y_{e,\Ni{56}}-Y_{e,\Fe{54}+\Ni{58}}}, 0 \right)
\end{equation}

where $Y_{e,i}$ is the electron fraction of substance $i$. The electron fraction in the fully burned material is
\begin{equation}
    Y_{e,n} = \frac{Y_e - Y_{e,f}(1-\phi_{qn})}{\phi_{qn}}
\end{equation}

and the electron fraction in the unburned fuel is
\begin{eqnarray}
    Y_{e,f} = X_{\C{12}}Y_{e,\C{12}} + X_{\Ne{20}}Y_{e,\Ne{20}} \\
    \nonumber + X_{\Ne{22}}Y_{e,\Ne{22}} + X_{\Ox{16}}Y_{e,\Ox{16}}
\end{eqnarray}

and the electron fractions for the remaining substances are calculated as usual, e.g. $Y_{e,\C{12}}$ = $Y_{e,\Ne{20}}$ = $Y_{e,\Ox{16}}$ = 0.5; $Y_{e,\Ne{22}} = \frac{10}{22} \sim 0.4545$; and $Y_{e,\Fe{54}+\Ni{58}} = \frac{1}{2}(\frac{26}{54} + \frac{28}{58}) \sim 0.48212$.

\section{Simulation and Results} \label{sec:results}
The evolution of the simulation is described in general in Section \ref{subsec:overview}, with explanations of the most noteworthy features and their connection to the physical system. This is followed by a description of how the final system is split into two parts \textemdash remnant and ejecta (Section \ref{subsec:separation}). The final yields are tabulated and discussed (Section \ref{subsec:yields}) alongside notable correlations (Section \ref{subsec:correlations}) and an analysis of the changing kinetic energy of the system (Section \ref{subsec:ke}). These results are then compared to a DDT simulation by \citet{Augustine_2019} that uses the same progenitor (Section \ref{subsec:DDT}), as well as to observational data (Section \ref{subsec:obs}).

\subsection{Overview of Explosion Stages} \label{subsec:overview}
\begin{figure*}[ht!]
    \plotone{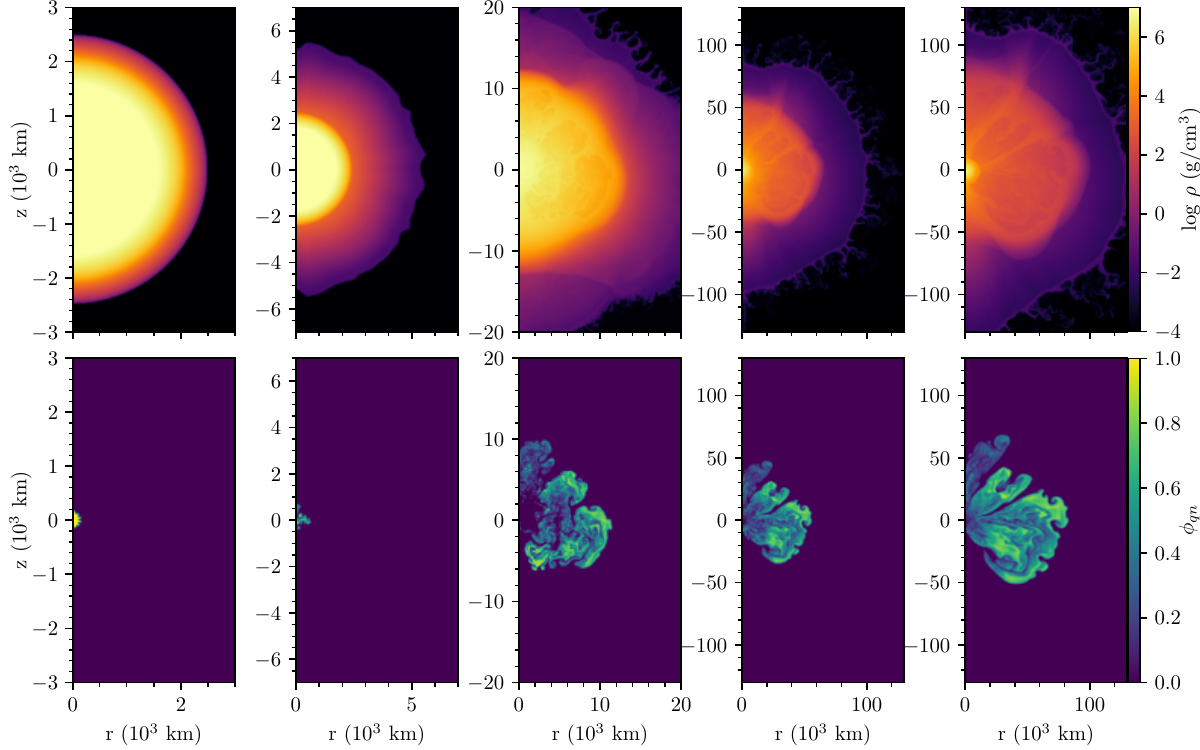}
    \caption{Density (top row) and $\phi_{qn}$ (bottom row) of R25 at 0, 0.7, 3, 13, and 20s of simulation time.}
    \label{fig:R25}
\end{figure*}

Figure \ref{fig:R25} shows both the density and percentage of material burned to IGE ($\phi_{qn}$) of an average realization, R25, at 0, 0.7, 3, 13, and 20s of simulation time.
Within the first second, the surface of the star puffs up and mixes with the fluff, creating artistic but entirely unphysical tendrils of swirling material. This effect is illustrated in the last three panels of Figure \ref{fig:R25}. The initial matchhead is put into pressure equilibrium with the surrounding star, but it is still a perturbation to the initial HSE as it introduces an early change in composition and density. The outward propagation of sound waves due to this perturbation is what we believe causes this sudden surface expansion.

In the meantime at the star’s center, the ADR scheme described in Section \ref{subsec:flame} starts to propagate the model flame, approximating the physical flame which heats the surrounding cold fuel by conduction until it begins to fuse. The resulting released nuclear energy is absorbed by the newly burned material as heat, increasing its temperature and causing it to expand, which decreases its density. This leaves pockets of hot, less dense material surrounded by cold, denser fuel. The Rayleigh-Taylor (RT) instability at this boundary perturbs the flame surface, causing plumes to form, merge, and wrinkle, which increases the surface area of the growing flame so it spreads faster. Our model captures the geometry of this RT instability at a larger resolved scale ($\Delta x = 4$ km), and adjusts the flame speed accordingly to account for unresolved small-scale motion. This strategy has known limitations due to the absence of realistic turbulence in 2D.

The plumes begin to rise and accelerate due to buoyancy, pushing the unburned fuel out of the way and imparting it with kinetic energy. The burned region is also increasing in size due to the flame propagating. The effect of these processes is an overall expansion of the star. As the flame front propagates and plumes rise, the density of the unburned fuel the flame propagates into decreases. A physical flame in a lower density environment would become slower and wider, since it takes longer to burn material at lower densities. At some point, the corresponding steady-state flame at a density of around $5 \times 10^7$ g/cm$^3$ would grow wider than the radius of the star. Our artificial flame is always 4 grid zones wide, so these effects are modeled by a decrease in the laminar flame speed and an increase in the evolution timescales of the progress variables $\phi_{aq}$ and $\phi_{qn}$. Eventually the timescales become so long that material essentially stops burning, a bit before 3s. After this point, there is no more energy being added to the system from fusion, and the expanding material becomes ballistic. Gravity will slow this expansion. Unlike DDT models, where the supersonic detonation produces a fairly uniform and rapid expansion, the system remains dynamic, and there are motions and interactions between the burned and unburned material.

The low carbon fraction in the star’s center means that less binding energy is released during fusion than in a regular C/O WD. This causes the flame to propagate more slowly, allowing the star more time to expand and leading to less buoyant plumes and more incomplete burning. Most of the star remains unburned, only now more mixed and less dense due to expansion. Because of this, most of the star remains behind and its gravitational pull causes most of the surrounding material to fall back, creating a remnant of unburned fuel mixed with ashes from burning. This fallback causes small shockwaves that push mass back out, but do not give it enough energy to fully escape the gravitational force, so that the mass falls back again and the process repeats. This behavior will hereafter be referred to as the pulsation of the remnant. Simultaneously, in contrast to this fallback, the puffed up surface of the star as well as the outer layers of the burned plumes have enough kinetic energy to escape the gravitational pull of the remnant, and travel out into space, continuing to decrease in both temperature and density.

\subsection{Stopping with Separated Ejecta and Remnant} \label{subsec:separation}
The system is separated into two pieces \textemdash ~remnant and ejecta. It is assumed that all the bound material will eventually fall back and become part of the remnant, while the ejecta will move off into space to be detected by our instruments on Earth. Therefore, the ejecta yields alone will be compared to yields calculated from observational data. We follow the evolution of this remnant-ejecta system out to 20s of simulation time, when the ejecta mass has become constant with time. 

Too much \edit1{farther in time} than this point, the ejecta and the fluff would begin to interact, and the fluff would decelerate the ejecta, imparting a \edit1{nonphysical} effect that would give an underestimate for the ejecta kinetic energy. This could be mitigated by implementing a density rolloff of the fluff. Without a detonation to shock the surface layers of the star, this becomes the biggest challenge to continuing this simulation to longer times.

Another issue with running these simulations out to long times is the disparate time evolution between the ejecta and the remnant. The ejecta is relatively cool, sparse, and settling into free expansion in a fixed form. In contrast, the remnant is hot and dense, collapsing in on itself and mixing on short timescales. This behavior is apparent in Figure \ref{fig:R25}; the expanding plumes remain essentially constant in shape from 13 s to 20 s, while the remnant is gaining mass and starting to pulsate. Resolving the simulation at the timescale of the remnant \edit1{ is a severe limitation on the evolution of the ejecta.} As our primary interest \edit1{for this 2D study} is in the ejecta yields, 20 s is long enough to obtain the desired information, but this issue will \edit1{require additional care when running future 3D simulations and creating} lightcurves and spectra, as discussed in Section \ref{sec:future}. 

Other studies \citep{Kromer_2015, Lach_2021} deresolve the remnant in order to observe the ejecta at later times, but even if we followed this prescription, our simulations would still be forced to stop short. While the majority of our realizations reach the full 20 s, some run into the region of parameter space described in Section \ref{subsec:initparams} that our EOS is not made to handle, even with the increased temperature floor. As we did not want to further increase the temperature floor and put more energy into our simulation, and as the ejected masses had leveled off already, we did not run these realizations further. All simulations made it past 15 s. The data reported in Section \ref{subsec:yields} are different quantities at the end of our simulation. It is important to note that these final yields refer to the yields at the end of each realization, which is either 20 s or at the point where the simulation stopped due to the above described EOS difficulty. The accompanying correlations between variables should not be affected by the difference in runtimes however, as the quantities of interest have leveled off, and should not change much over time.

\subsection{Final Yields} \label{subsec:yields}
\begin{deluxetable*}{llccc}[ht!] \label{tab:yields}
\tablecaption{The range and interquartile range (IQR, middle 50\%) of the final yields from our simulation suite.}
\tablehead{
\colhead{Calculated Quantity} & & \colhead{Total} & \colhead{Bound} & \colhead{Unbound}
}
\startdata
Mass (\Msun) & Range & 1.36 & 1.230 - 1.351 & 0.012 - 0.133 \\
             & Middle 50\% & 1.36 & 1.289 - 1.332 & 0.031 - 0.074 \\
\hline
Estimated \Ni{56} (\Msun) & Range & 0.066 - 0.127 & 0.064 - 0.115 & 0.00030 - 0.01528 \\
                          & Middle 50\% & 0.095 - 0.110 & 0.090 - 0.100 & 0.00187 - 0.00738 \\
\hline
IGE (\Msun) & Range & 0.19154 - 0.28051 & 0.18288 - 0.26143 & 0.00117 - 0.04274 \\
            & Middle 50\% & 0.22135 - 0.25203 & 0.20828 - 0.23183 & 0.00614 - 0.01833\\
\hline
NSQE (\Msun) & Range & 0.03286 - 0.05497 & 0.03203 - 0.5497 & 0.00017 - 0.00614 \\
             & Middle 50\% & 0.04600 - 0.05302 & 0.04454 - 0.04889 & 0.00087 - 0.00298\\
\hline
Ash (\Msun) & Range & 0.09079 - 0.16329 & 0.08939 - 0.16149 & 0.00033 - 0.01227\\
            & Middle 50\% & 0.11888 - 0.13720 & 0.11610 - 0.13252 & 0.00162 - 0.00647\\
\hline
Fuel (\Msun) & Range & 0.90380 - 1.04757 & 0.82241 - 1.01695 & 0.00890 - 0.10458 \\
             & Middle 50\% & 0.92959 - 0.96152 & 0.87219 - 0.94086 & 0.01783 - 0.05095 \\
\hline
\Ni{56} : IGE Ratio & Range & 0.344 - 0.471 & 0.350 - 0.499 & 0.161 - 0.476\\
                    & Middle 50\% & 0.402 - 0.441 & 0.408 - 0.449 & 0.297 - 0.406\\
\hline
\Ni{56} : Mass Ratio & Range & 0.04848 - 0.09305 & 0.04842 - 0.09274 & 0.02428 - 0.12240 \\
                     & Middle 50\% & 0.0699 - 0.0809 & 0.06880 - 0.08024 & 0.05695 - 0.09506\\
\hline
KE (10$^{49}$ergs) & Range & 0.89597 - 3.46294 & 0.45480 - 2.29050 & 0.10703 - 1.66041\\
                    & Middle 50\% & 1.6212 - 2.3807 & 0.9982 - 1.5614 & 0.3445 - 0.8257
\enddata
\tablecomments{Bound and unbound material are distinguished based on the value of $E_{\rm binding}$ (defined by Equation \ref{eq:BE}) having a negative or positive value, respectively. The IGE, NSQE, Ash, and Fuel yields correspond to $\mathcal{X}_N$, $\mathcal{X}_q$, $\mathcal{X}_a$, and $\mathcal{X}_f$ described in \edit1{Appendix \ref{ap:flame}} and Section \ref{subsec:flame}, and the estimate of \Ni{56} was calculated using Equation \ref{eq:Ni}. It is important to note that the minimum values listed do not all come from the same realization (and the same with the maximum values).}
\end{deluxetable*}

\begin{figure}[ht!]
    \includegraphics[]{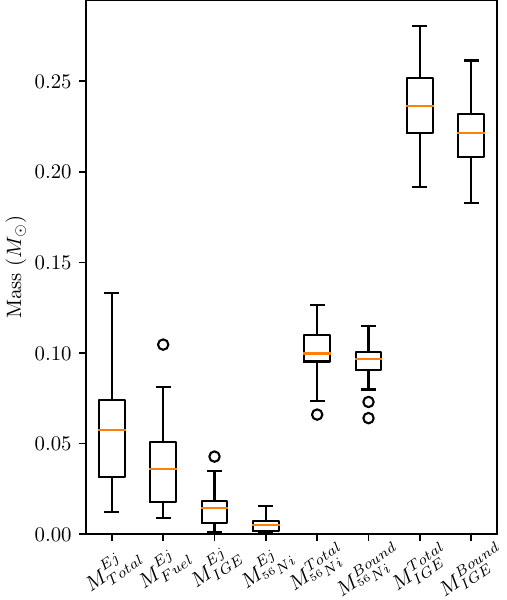}
    \caption{A visual representation of selected final integrated quantities from Table \ref{tab:yields}. Shown are box-and-whisker plots of (from left to right) the total ejected mass ($M^{\rm{Ej}}_{\rm{Total}}$), the ejected fuel ($M^{\rm{Ej}}_{\rm{Fuel}}$), the ejected IGE mass ($M^{\rm{Ej}}_{\rm{IGE}}$), the ejected \Ni{56} mass ($M^{\rm{Ej}}_{\Ni{56}}$), the total \Ni{56} mass ($M^{\rm{Total}}_{\Ni{56}}$), the bound \Ni{56} mass ($M^{\rm{Ej}}_{\Ni{56}}$)the total IGE mass ($M^{\rm{Total}}_{\rm{IGE}}$), and the bound IGE mass ($M^{\rm{Bound}}_{\rm{IGE}}$). \edit1{The individual points are data that lie outside the whisker range,  1.5 $\times$ IQR away from the 75th and 25th percentiles.} Most of the star remains bound, and the majority of the ejecta is unburned fuel. Most of the produced $\Ni{56}$ and IGE remains bound.}
    \label{fig:Boxplot}
\end{figure}

\vspace{-30pt}
The ranges of ejecta and remnant yields and energies are shown in Table \ref{tab:yields}, visually supplemented by the box-and-whisker in Figure \ref{fig:Boxplot}. 
The table reports the full range and also the middle 50\% (the interquartile range, or IQR) of the data. Each box-and-whisker plot shows the median (50th percentile, orange line) of each quantity, with upper and lower box boundaries marking their 75th and 25th percentiles, respectively. The range enclosed by the box corresponds to the IQR of the data, which is also reported in Table \ref{tab:yields}. Note that the median of the data is not necessarily equidistant from the 25th and 75th percentiles (ie not necessarily in the center of the box). The whiskers are plotted at 1.5 $\times$ IQR away from the 75th and 25th percentiles, but do not extend beyond the limits of the data. Data that lie outside the range of the whiskers are plotted as individual points.

Notably, $\sim$ 66 - 77\% of the star remains unburned; $\sim$ 14 - 21\% is fully burned to IGE; only $\sim$ 2 - 4\% is partially burned to NSQE; and $\sim$ 7 - 12\% undergoes carbon burning and then stops (the ash state). Only $\sim$ 5 - 9\% of the star is burned to \Ni{56}, which is $\sim$ 40\% of the IGE. In addition, the vast majority, $\sim$ 90 - 99 \%, of the star remains bound, demonstrating that a deflagration does indeed produce much smaller amounts of ejected material and high-mass elements than a DDT. The small amount of remaining mass has enough energy to overcome the gravitational pull of the remnant, and is ejected from the system. As demonstrated in Figure \ref{fig:Boxplot}, most of this ejected material is unburned fuel, but it also contains the edges of the plumes of burned material. Interestingly, the majority of the burned material, including the IGE and \Ni{56}, is bound, and will fall back and join the remnant.

\begin{figure*}[ht!]
\gridline{\leftfig{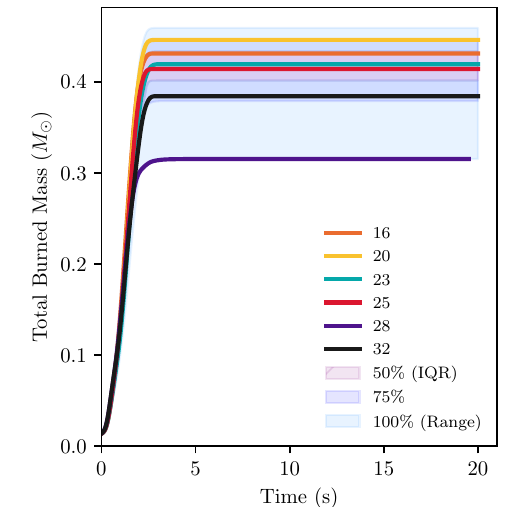}{\columnwidth}{(a)}
          \rightfig{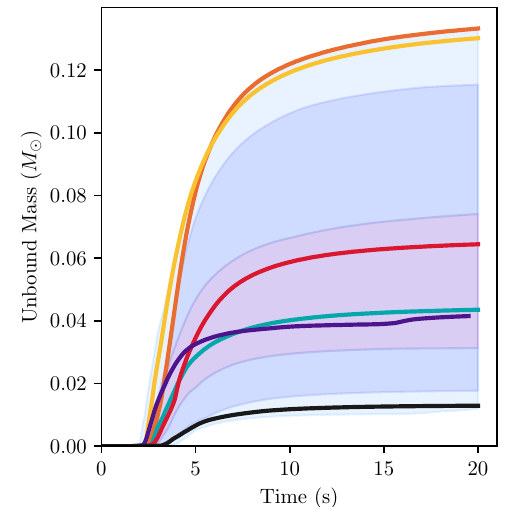}{\columnwidth}{(b)}}
\gridline{\leftfig{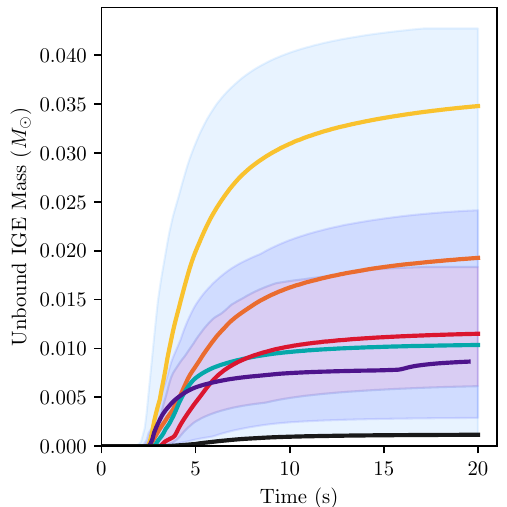}{\columnwidth}{(c)}
          \rightfig{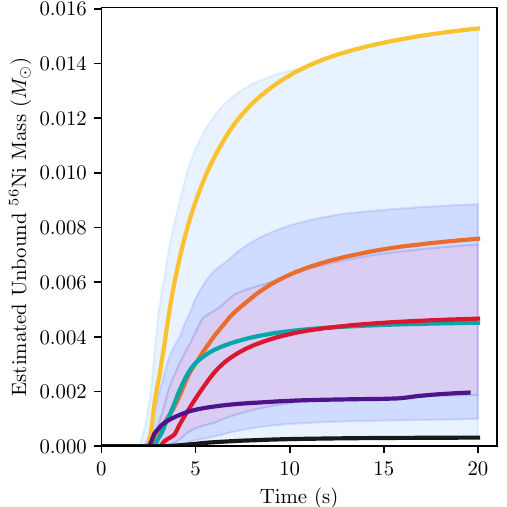}{\columnwidth}{(d)}}
    \caption{Plots over time of (a) the total burned mass, (b) unbound mass, (c) unbound IGE mass, and (d) unbound \Ni{56} mass. The results of the following six simulations are highlighted: R16 (orange), which produced the most ejected mass; R20 (yellow) and R32 (black), which yielded the most and least ejected \Ni{56} mass, respectively; R28 (purple), which by far burned the least mass; and R23 (blue) and R25 (fuchsia), two average runs that produced almost the same ejected \Ni{56} mass but exhibit different rise times. Colors are in dark to light ordered by ejected \Ni{56} mass. To get a sense of the range and localization of the data, also shown are the IQR for each timestep (purple, shaded), the range of the middle 75\% of the data (blue, shaded), and the full range of the data (light blue, shaded).}
    \label{fig:Time_IQR}
\end{figure*}

One important feature to note is that the simulation that produced the largest ejected mass (R16) is neither the one corresponding to the most ejected \Ni{56} mass (R20), nor the one with the most ejected IGE (R14), nor the one that burned the most mass (R19). Similarly, the simulation that produced the least ejected \Ni{56} and ejected IGE (R32) has neither the least ejected mass (R35) nor the least burned mass (R28). This feature is more clearly demonstrated in Figure \ref{fig:Time_IQR}, which shows how these quantities evolve with time.

In these plots, the results of the following six simulations are highlighted: R16, which produced the most ejected mass; R20 and R32, which yielded the most and least ejected \Ni{56} mass, respectively; R28, which by far burned the least mass; and R23 and R25, two average runs that produced almost the same ejected \Ni{56} mass but exhibit different rise times. Highlighted in each plot is the full range of the data from the suite, and also shown is the region where 75\% of the data is distributed, and the region where 50\% of the data is distributed (the IQR). These measures were chosen rather than the standard deviation, because they give a better picture of how our realizations are distributed. The results are not perfectly Gaussian, and have a large range; therefore the standard deviation is often greater than the average, and shading in that region would give a misleading representation of our data. Therefore, we use quartiles as a better description of our data distribution. It can be seen that the full range of the data is about twice as large as the middle 50\% of the data.

In general, the more mass that is burned by the flame, the more material that is converted to IGE, and the more nuclear energy that is released, which leads to higher kinetic energies imparted to the surrounding star, and therefore more material that can break free from the remnant and is ejected. While this reasoning seems straightforward, things are complicated by both the nonlinearity of this problem, especially the RT instability, and the overly restrictive cylindrically-symmetric 2D geometry. The RT instability determines how the initial perturbations evolve into plumes of burned material, which plumes overtake others and which become dominant, and how the plumes combine, deciding what parts of the star will contain the most burned material. It also decides the initial direction that these plumes move, which means they are either headed towards the less dense, colder material towards the edge of the star, or towards the hot dense fuel of the inner star, which decides how long the plumes can burn.

\begin{figure}[ht!]
    \plotone{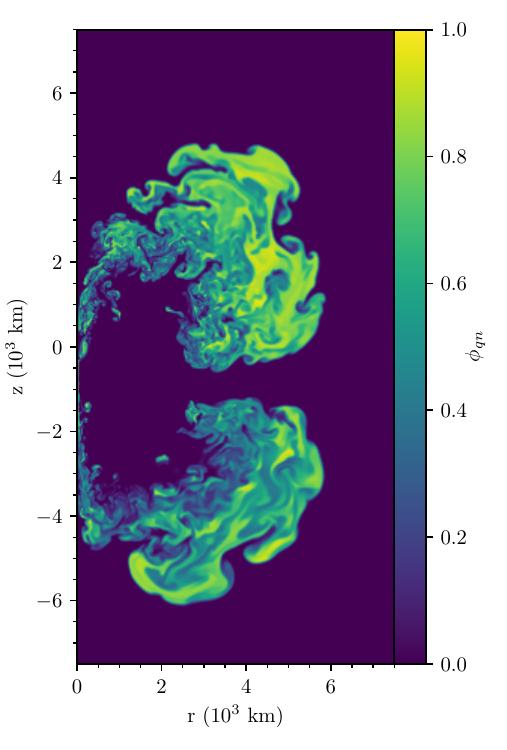}
    \caption{Fraction of material burned to IGE ($\phi_{qn}$) of R28 at 2.0s of simulation time. The curvature of the plumes due to the cylindrical geometry of the setup is apparent.}
    \label{fig:curvature}
\end{figure}

Our choice of geometry, on the other hand, causes the plumes to move away from the z-axis and curl around the mostly unburned core of the star. In our cylindrically symmetric geometry, each grid cell corresponds to an annulus of material, which means that grid cells along the r-axis occupy a larger volume and contain more mass than those along the z-axis. This means that plumes parallel to the z-axis tend to rise faster and burn through less mass, since they have less material overhead. This creates a shear vortex that rolls the plume heads towards the r-axis \citep{Townsley_2009}. As discussed in Section \ref{subsec:initflame} the initial flame morphologies are explicitly chosen to not have a radial maximum on the z-axis axis because that direction has effectively too little resistance to plume rise due to the imposed symmetry -- allowing a maximum there leads to excessively unrealistic pathologies. This curving is apparent in R28, shown in Figure \ref{fig:curvature}, which has the least total burned mass and the least total IGE mass, but ejects an amount of mass and IGE within the middle 50\% of all the realizations.

\edit1{\subsection{Sensitivity to Initially Burned Region}} \label{subsec:initial}
\edit1{It is important to quantify the effect that the shape and size of the initially burned region (matchhead) has on the results, 
as the matchhead is only necessary because the realistic initial flame scale cannot be resolved. As seen in Section \ref{subsec:initflame}, the matchhead is described by Equation \ref{eq:ignition}, where each realization has a different set of $A_\ell$.}

\begin{figure*}[ht!]
\gridline{\leftfig{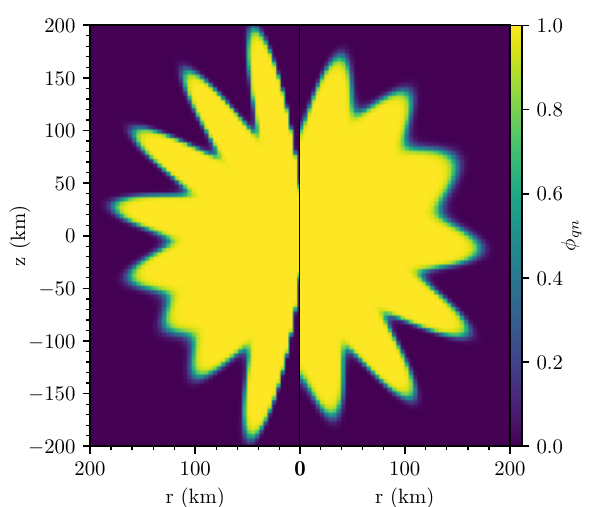}{\columnwidth}{(a)}
          \rightfig{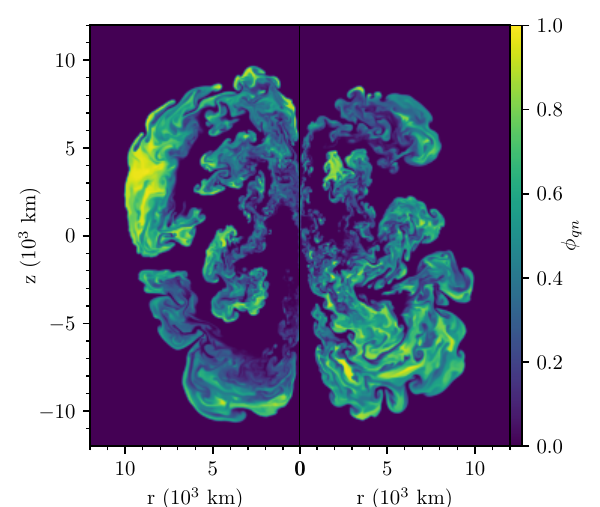}{\columnwidth}{(b)}}
          \vspace{-13pt}
\gridline{\leftfig{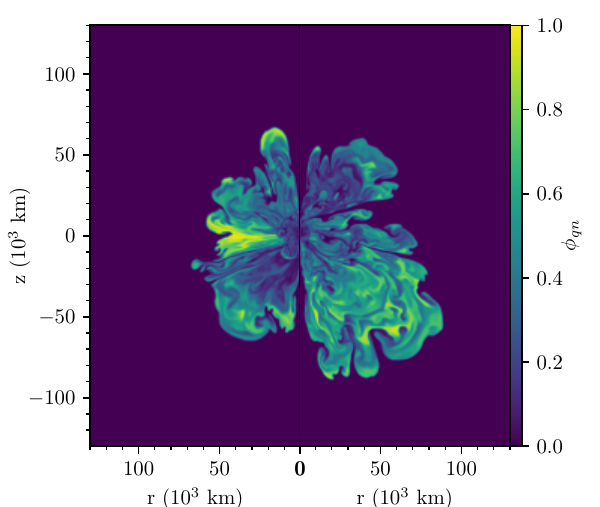}{\columnwidth}{(c)}
          \rightfig{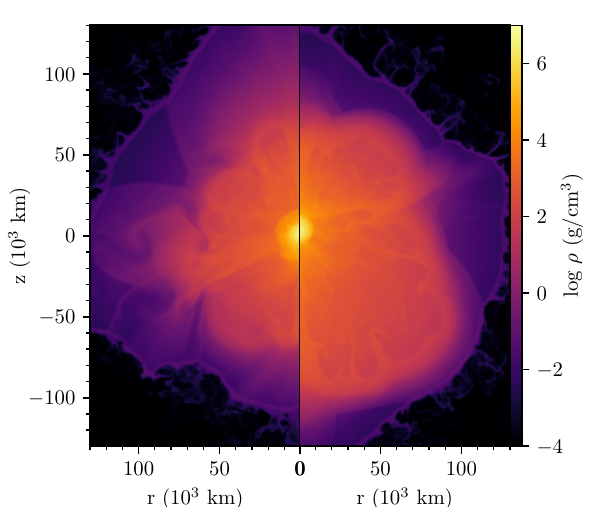}{\columnwidth}{(d)}}
    \caption{A comparison of R32 (left panels, least ejected \Ni{56} mass) and R20 (right panels, most ejected \Ni{56} mass) over time. The (a) initial flame, (b) the  the fraction of material burned to IGE at 3s, (c) the fraction of material burned to IGE at 20s, and (d) the density at 20s, are shown.}
    \label{fig:2realz}
\end{figure*}

\begin{figure*}[ht!]
\gridline{\leftfig{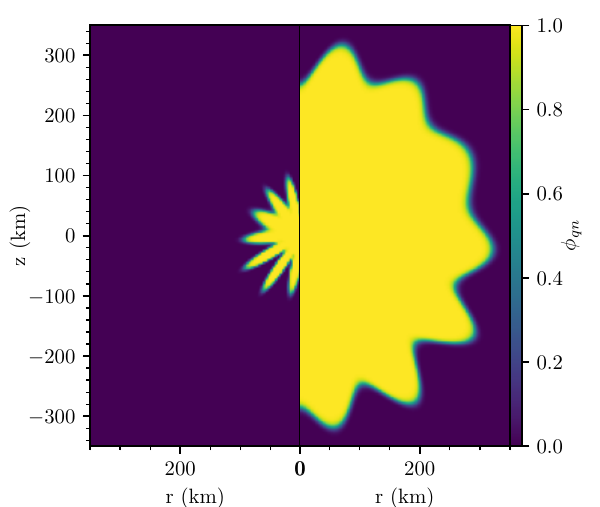}{\columnwidth}{(a)}
          \rightfig{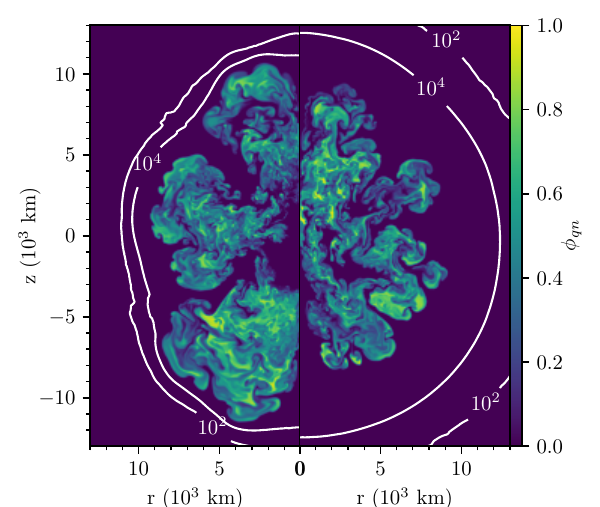}{\columnwidth}{(b)}}
          \vspace{-13pt}
\gridline{\leftfig{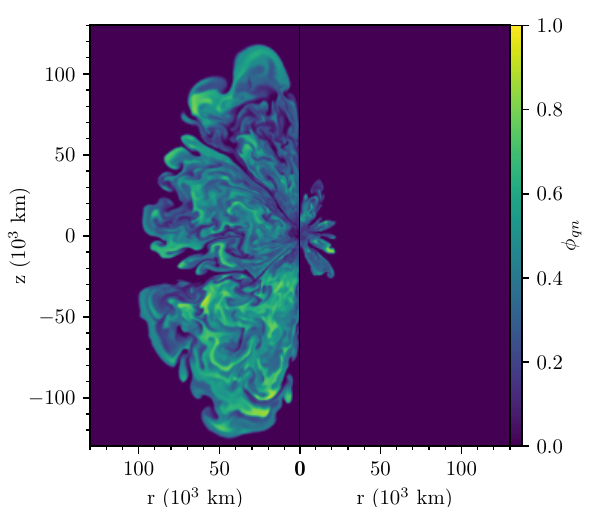}{\columnwidth}{(c)}
  \rightfig{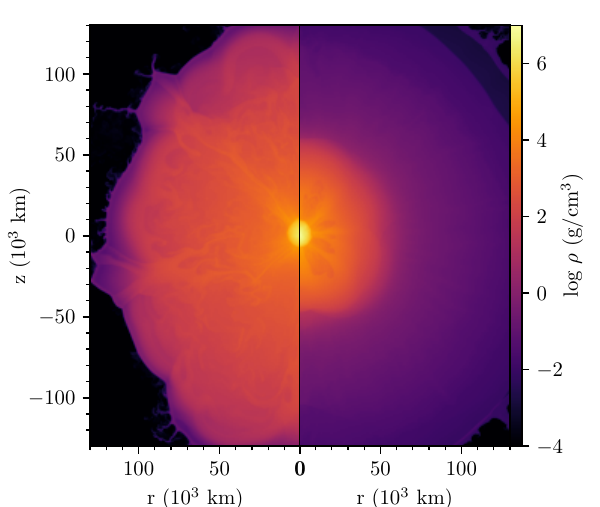}{\columnwidth}{(d)}}
    \caption{A comparison of the half radius (left panels) and double radius (right panels) matchhead sizes for R20 over time. The (a) initial flame, (b) the  the fraction of material burned to IGE at 3s, (c) the  the fraction of material burned to IGE at 20s, and (d) the density at 20s, are shown. In (b), density contours at 10$^2$ and 10$^4$ g/cm$^3$ are also plotted.}
    \label{fig:2realz_initFlame}
\end{figure*}

Due to the nonlinearity of this problem, it is nearly impossible to relate the shape of the matchhead to the ejecta masses. The resulting ejecta mass is strongly correlated with neither the initial burned mass (a measure of the volume of mass inside the flame at t=0), nor the maximum or minimum radius the flame plumes reach. Figure \ref{fig:2realz} shows the realization with the least ejected \Ni{56} (R32) side-by-side with the realization that produced the most (R20). The burned regions look similar at 3s, but R32’s main plume is curling inward, whereas R20’s plumes are all moving outward. 

This difference shows that the early flame evolution can create large differences in outcome, but it is difficult to predict exactly which initial parameters and in what combination produce the effect.
This \edit1{lack of correlation between the initial flame geometry and the final yields} shows that the choice of initial conditions for the simulation suite (discussed in Section \ref{subsec:initflame}) successfully creates a random sample \edit1{in the sense that the shape of the initial flame cannot be directly related to the yields}.

\edit1{However, changing the radius of the initial matchhead does produce a systematic effect on the results. From our main suite of 30 realizations, we selected 10 that spanned the full range of ejected \Ni{56} yields.
We investigated how doubling (300 km) and halving (75 km) the matchhead radius impacted the final yields, using the same prescription as before in Section \ref{sec:analysis} to analyze the results.}

\begin{figure*}[ht!]
\gridline{\leftfig{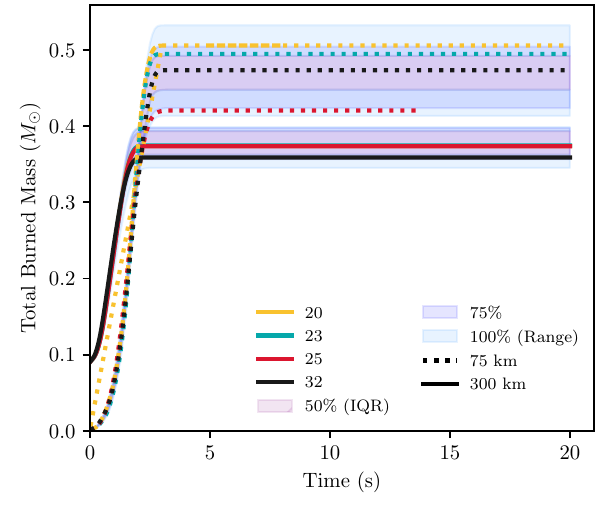}{\columnwidth}{(a)}
\rightfig{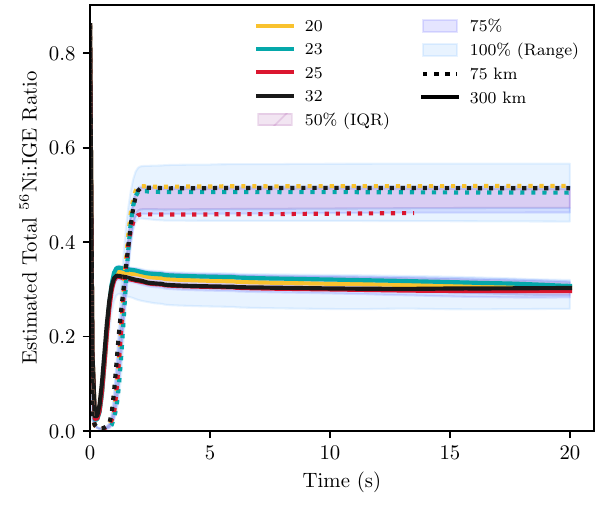}{\columnwidth}{(b)}}
\gridline{\leftfig{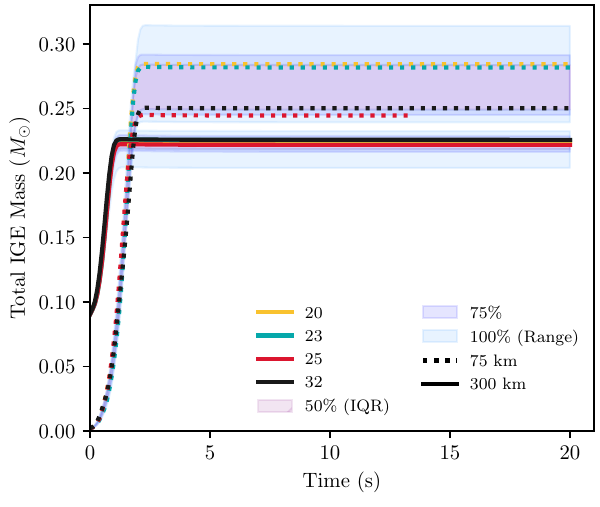}{\columnwidth}{(c)}   \rightfig{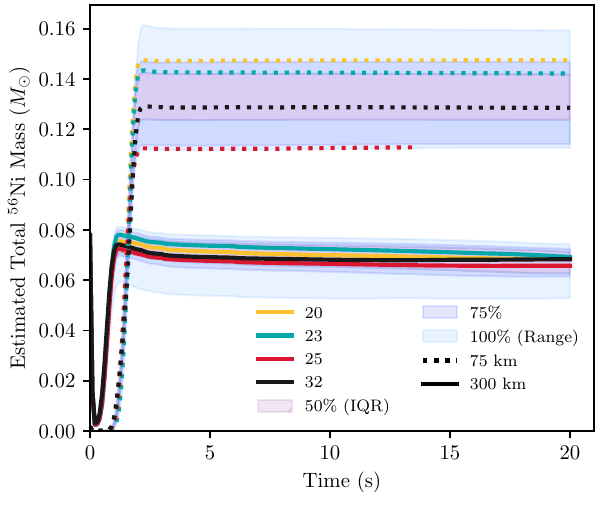}{\columnwidth}{(d)}}
    \caption{Plots over time of (a) the total burned mass, (b) total \Ni{56} : IGE Ratio, (c) total IGE mass, and (d) total \Ni{56} mass for the 75 km (dotted lines) and 300 km (solid lines) radius matchheads. Similarly to Figure \ref{fig:Time_IQR}, the results of the following four simulations are highlighted: R20 (yellow) and R32 (black), which yielded the most and least ejected \Ni{56} mass for the baseline (150 km) radius matchhead, respectively; and R23 (blue) and R25 (fuchsia), two average runs that produced almost the same ejected \Ni{56} mass but exhibit different rise times for the baseline (150 km) radius matchhead. Also shown are the IQR for each timestep (purple, shaded), the range of the middle 75\% of the data (blue, shaded), and the full range of the data (light blue, shaded).}
    \label{fig:iax_sidebyside}
\end{figure*}

\begin{deluxetable*}{llccc}[ht!] \label{tab:HalfIaxData}
\tablecaption{The range and interquartile range (IQR, middle 50\%) of the final yields from the halved (75 km) matchhead radius simulation suite.}
\tablehead{
\colhead{Calculated Quantity} & & \colhead{Total} & \colhead{Bound} & \colhead{Unbound}
}
\startdata
Mass (\Msun) & Range & 1.36 & 0.993 - 1.296 & 0.069 - 0.372 \\
            & Middle 50\% & 1.36 & 1.124 - 1.268 & 0.097 - 0.240 \\
\hline
Estimated \Ni{56} (\Msun) & Range & 0.113 - 0.159 & 0.094 - 0.126 & 0.007 - 0.040 \\
                          & Middle 50\% & 0.124 - 0.142 & 0.108 - 0.113 & 0.016 - 0.032 \\
\hline
IGE (\Msun) & Range & 0.239 - 0.314 & 0.190 - 0.246 & 0.016 - 0.079 \\
            & Middle 50\% & 0.249 - 0.284 & 0.206 - 0.236 & 0.037 - 0.059\\
\hline
\Ni{56} : IGE Ratio & Range & 0.444 - 0.566 & 0.463 - 0.571 & 0.344 - 0.551 \\
            & Middle 50\% & 0.472 - 0.512 & 0.472 - 0.517 & 0.441 - 0.499\\
\hline
\Ni{56} : Mass Ratio & Range & 0.083 - 0.117 & 0.074 - 0.127 & 0.079 - 0.190 \\
            & Middle 50\% & 0.091 - 0.104 & 0.089 - 0.099 & 0.105 - 0.157\\
\hline
KE (10$^{49}$ergs) & Range & 1.901 - 6.041 & 0.867 - 2.119 & 0.711 - 4.666\\
                    & Middle 50\% & 2.624 - 4.520 & 1.114 - 1.515 & 1.552 - 3.193 \\
\enddata
\end{deluxetable*}

\begin{deluxetable*}{llccc}[ht!] \label{tab:DoubleIaxData}
\tablecaption{The range and interquartile range (IQR, middle 50\%) of the final yields from the doubled (300 km) matchhead radius simulation suite.}
\tablehead{
\colhead{Calculated Quantity} & & \colhead{Total} & \colhead{Bound} & \colhead{Unbound}
}
\startdata
Mass (\Msun) & Range & 1.36 & 1.32815 - 1.34920 & 0.00048 - 0.02142 \\
            & Middle 50\% & 1.36 & 1.34870 - 1.34895 & 0.00066 - 0.00092 \\
\hline
Estimated \Ni{56} (\Msun) & Range & 0.053 - 0.074 & 0.053 - 0.074 & 0.000 - 0.002 \\
                          & Middle 50\% & 0.063 - 0.071 & 0.063 - 0.071 & 0.000 - 0.000 \\
\hline
IGE (\Msun) & Range & 0.204 - 0.232 & 0.204 - 0.229 & 0.000 - 0.004 \\
            & Middle 50\% & 0.218 - 0.226 & 0.218 - 0.226 & 0.000 - 0.000 \\
\hline
\Ni{56} : IGE Ratio & Range & 0.259 - 0.320 & 0.259 - 0.319 & 0.327 - 0.353 \\
            & Middle 50\% & 0.288 - 0.314 & 0.288 - 0.314 & 0.333 - 0.346\\
\hline
\Ni{56} : Mass Ratio & Range & 0.039 - 0.054  & 0.039 - 0.055 & 0.000 - 0.071 \\
            & Middle 50\% & 0.046 - 0.052 & 0.046 - 0.053 & 0.000 - 0.000\\
\hline
KE (10$^{49}$ergs) & Range & 1.197 - 3.589 & 1.189 - 3.581 & 0.008 - 0.220 \\
                    & Middle 50\% & 2.239 - 3.017 & 2.227 - 2.938 & 0.011 - 0.014 \\
\enddata
\tablecomments{Eight of the ten realizations produced no ejected mass. The two that did produced very little, most of which is material from the expanded, unburnt star.}
\end{deluxetable*}

\vspace{-30pt}
\edit1{Examination of this data yields multiple interesting results. A spatial comparison of the results of these two matchhead radii for the same realization (R20) can be seen in Figure \ref{fig:2realz_initFlame}.} 
\edit1{Perhaps the most prominent feature is that the larger matchead produces much smaller and more constrained plumes of burned material. This is because the larger matchhead deposits much more energy into the star's center at the start of the simulation, causing the star to expand faster. Additionally, the larger matchhead suppresses the flame instabilities that boost the burning; the smaller matchhead's plumes will evolve in the dense center of the star subject to the RT instability, while the larger matchhead has a flat profile in the denser region. Due to this expansion and suppression, a larger radius matchhead burns through less dense material more slowly, causing the burning to extinguish sooner and less material to be burned to IGE than for a smaller matchhead. This means the buoyant plumes of burned material will not only be smaller in size, but also will rise less because of the lower density of material surrounding them, leading to a smaller overall KE. This can be seen in Figure \ref{fig:2realz_initFlame}; the smaller matchhead's plumes reach to the edge of the expanding star, while the larger matchhead's plumes are much more constrained.}

\edit1{As shown in Figure \ref{fig:iax_sidebyside}, as the initial matchhead radius increases, the total kinetic energy, \Ni{56}, and IGE yields decrease, as does the \Ni{56}:IGE ratio. Additionally, the burning stops sooner.} 
\edit1{The final yields from these two different matchhead radii are shown in Tables \ref{tab:HalfIaxData} and \ref{tab:DoubleIaxData}. Most of these yields have some overlap with those for the 150 km radius matchhead (see Table \ref{tab:yields}) discussed in the rest of this work, with the 75 km matchhead overlapping the higher range, and the 150 km matchhead overlapping the lower range. An important observation from this data is that eight of the ten double matchhead radius realizations produced no ejected mass. In these instances, all of the mass remained gravitationally bound and fell back towards the progenitor, suggesting that the white dwarf will re-form without any net mass loss.}

\edit1{It is important to note that all of these initial flames, regardless of size, are but an approximation, as a realistic flame is smaller than the size of a grid cell and therefore cannot be resolved. Changing the initial size of the matchhead can be thought of as starting the simulation at a different point in time; a realistic flame would need less time to grow to the size of the smaller matchhead than the larger matchhead. The larger matchhead also expands the star faster and has suppressed flame instabilities. Logically, a smaller initial flame seems like it would be more realistic, but it is uncertain whether the ignition would be in one location and grow, or from several locations and then merge. Other studies test the differences between this by performing a similar initial conditions study with multiple ignition points and sizes, but this is beyond the scope of this work. A flame in 2D like ours cannot be turbulent, so is not realistic. Therefore, this test cannot say anything about the ``realism'' of the initial parameters of the flame, but rather serves as a measure of uncertainty for our simulation. Though the initial matchhead size produces a systematic effect, the yields remain in the suggested range of Type Iax events.}

\subsection{Correlations} \label{subsec:correlations}
\begin{figure*}[ht!]
    \includegraphics[angle=90, scale=0.8]{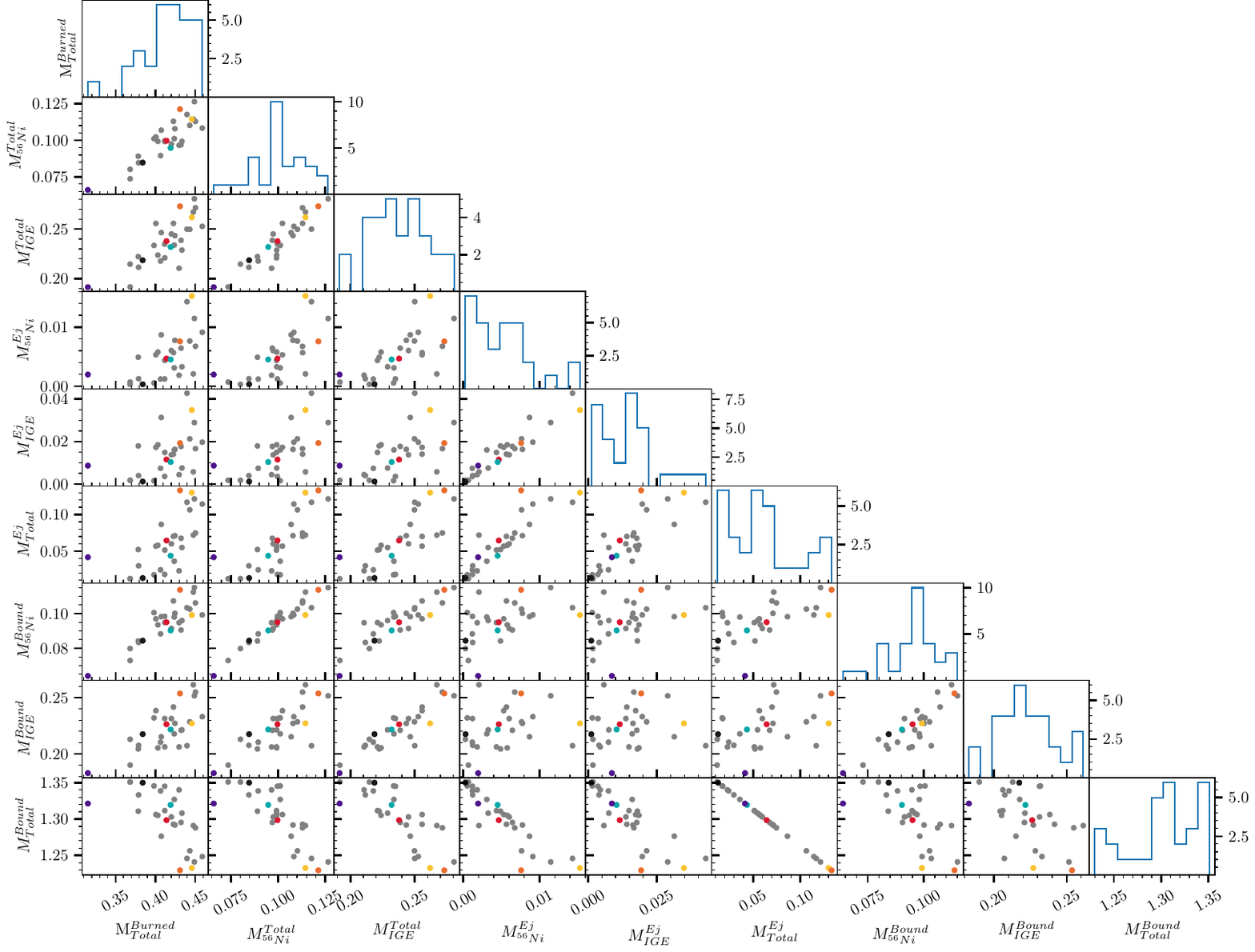}
    \caption{Corner plot showing correlations between different final integrated quantities \edit1{and their distributions}. All masses are in units of \Msun. Shown are the total burned mass ($M^{\rm{Burned}}_{\rm{Total}}$), the total \Ni{56} mass ($M^{\rm{Total}}_{\Ni{56}}$), the total IGE mass ($M^{\rm{Total}}_{\rm{IGE}}$), the ejected \Ni{56} mass ($M^{\rm{Ej}}_{\Ni{56}}$), the ejected IGE mass ($M^{\rm{Ej}}_{\rm{IGE}}$), the total ejected mass ($M^{\rm{Ej}}_{\rm{Total}}$), the bound \Ni{56} mass ($M^{\rm{Bound}}_{\Ni{56}}$), the bound IGE mass ($M^{\rm{Bound}}_{\rm{IGE}}$), and the total bound mass ($M^{\rm{Bound}}_{\rm{Total}}$).}
    \label{fig:corner_plot}
\end{figure*}

\edit1{Returning now to our suite of 30 realizations, the nonlinearity and curvature of the simulation} is one of many reasons why, even though there is the expected correlations between the burned mass, the ejected mass, and the IGE mass, they exhibit varying degrees of tightness and scatter. These correlations, among others, are shown in Figure \ref{fig:corner_plot}. The burned mass is made up of ash, NSQE elements, and IGE which includes \Ni{56}; therefore it has a positive correlation which each of these quantities, as shown in the first column of Figure \ref{fig:corner_plot}. These correlations display a scatter which generally increases with increases burned mass. Additionally, the more mass that is burned, the less total mass remains bound, though this is not a strong correlation. The IGE and \Ni{56} show a strong positive correlation, meaning that a reasonably well-defined \Ni{56}:IGE ratio can be calculated. This ratio is $\sim$ 0.4 for total, bound, and unbound quantities, but has a much larger range for the unbound quantities, as seen in Table \ref{tab:yields}. The perfect correlation between the total bound mass and the total unbound mass is due to the fact that the bound mass plus the unbound amss of the system must sum to the total mass, 1.36 \Msun. Unbound quantities tend to exhibit a stronger correlation than \edit1{bound} quantities, for example the graph of unbound \Ni{56} and unbound IGE is more strongly correlated than that of bound \Ni{56} and bound IGE. However, the unbound quantities tend to increase their scatter with increasing burned mass much more so than their bound counterparts. The quantity that exhibits the strongest determinant as to how much material will be bound vs unbound is the system's kinetic energy.

\subsection{Kinetic Energy} \label{subsec:ke}

\begin{figure}[ht!]
    \includegraphics[width=\columnwidth]{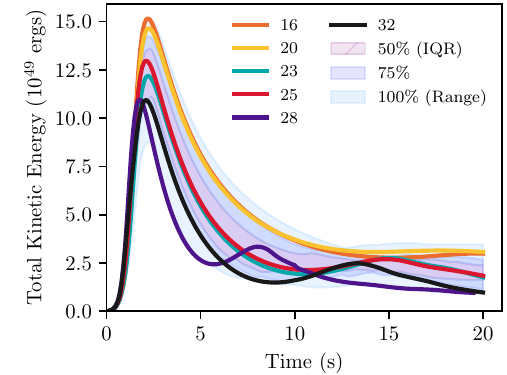}
    \caption{The total KE vs time, with the same feature highlights as Figure \ref{fig:Time_IQR}. The total KE peaks right as the bu ends, and then decreases as the gravitational pull of the remnant slows down the star's expansion. The KE peaks again once the remnant starts contracting and material falls back on it, and then decreases again, showing the pulsation of the remnant.}
    \label{fig:KE_Total_IQR}
\end{figure}

The pulsation of the remnant can be seen in the maximum density vs time as well as the total KE vs time. The latter is plotted in Figure \ref{fig:KE_Total_IQR}.


\begin{figure}[ht!]
    \includegraphics[width=0.965\columnwidth]{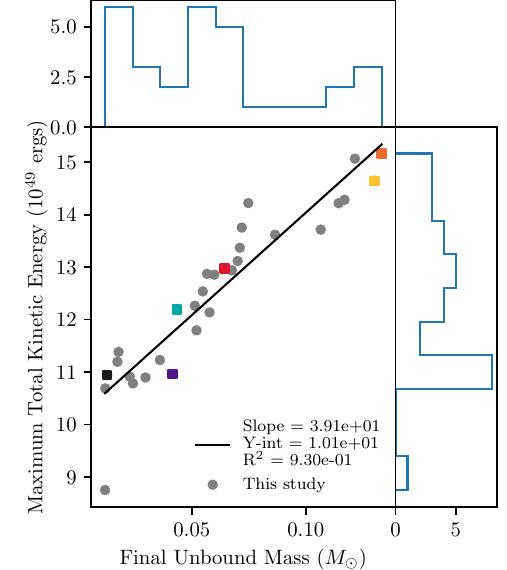}
    \includegraphics[width=0.965\columnwidth]{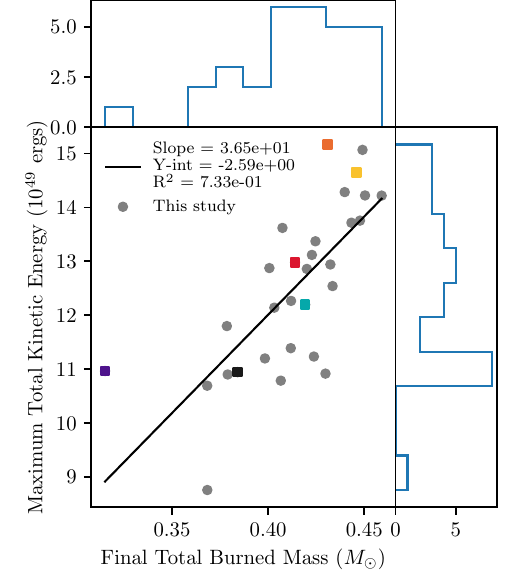}
    \caption{The maximum total kinetic energy of each simulation is plotted against both (top) the final unbound mass, and (bottom) the final total burned mass. The tight correlation between the former indicates that the maximum total KE determines how much mass can escape the gravitational pull of the remnant. The colored points show the same feature highlights as Figure \ref{fig:Time_IQR}.}
    \label{fig:KE_Total}
\end{figure}

Both quantities exhibit a damped oscillatory pattern. The maximum density quickly drops two orders of magnitude during burning, as the energy deposited by fusion expands the burned regions and pushes the rest of the star out with it. The densest part of the simulation is still the expanded core of the star. In general, realizations with less burned mass will reach higher densities faster, because the star is less disrupted by the burning, less expanded, and can therefore recombine into a remnant faster. Realizations that burned less mass also impart less KE, so they reach a smaller peak KE sooner. This picture is complicated by the fact that the KE is not evenly distributed throughout the star, and some simulations are left with more bound mass than others. R35 reaches the lowest peak KE and it has the most bound mass, whereas R13 reaches the highest peak KE but R16 has the least bound mass.

As shown in Figure \ref{fig:KE_Total_IQR}, the total KE reaches its maximum right around the time burning stops, since the energy released by fusion is what causes the temperature of the burned regions to increase, leading to expansion of the plumes that push mass out from the center of the star due to buoyancy. As soon as the burning stops, the total KE of the simulation starts to decrease, as it becomes converted into gravitational PE.

In general, more burned mass means more energy released and therefore a higher maximum KE. However, as shown in the bottom panel of Figure \ref{fig:KE_Total}, the amount of burned mass alone is not enough to determine the maximum KE of each realization. While there is a correlation between these two quantities, there is scatter, indicating that there are other factors at play. Due to the nonlinearity of the simulation, realizations that have a certain amount of burned mass will impart different amounts of KE depending on the location and direction of the plumes. This is also why the final burned mass is not strongly correlated with the ejecta yields. After the burning has stopped, however, the nonlinear effects from the RT instability and curving due to the geometry are not as prominent. In general, the kinetic energy imparted during the burning phase determines how much mass will be able to escape the gravitational pull of the remnant. Therefore, the maximum total KE is tightly correlated with the
\begin{figure}[ht!]
    \includegraphics[width=\columnwidth]{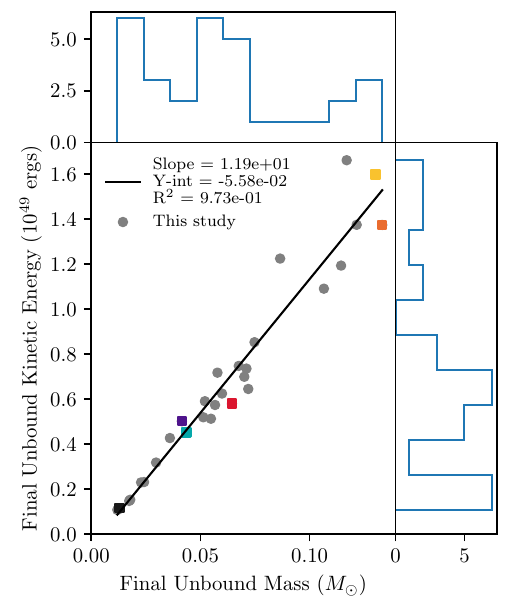}
    \caption{Total unbound kinetic energy (in $10^{49}$ ergs) vs the total unbound mass (in \Msun) of each realization. The colored points show the same feature highlights as Figure \ref{fig:Time_IQR}.}
    \label{fig:char_vel}
\end{figure}final unbound mass, as seen in the top panel of Figure \ref{fig:KE_Total}.

There appears to be a ``characteristic'' ejecta velocity, since the ejecta KE and ejected mass show a linear relationship, as seen in Figure \ref{fig:char_vel}. Using the slope of the line of best fit, this characteristic ejecta velocity can be estimated as 3500 km/s.

\subsection{Comparison to DDT model} \label{subsec:DDT}
The initial white dwarf progenitor model, initial flames, burning model, and simulation environment (FLASH) are identical to those in \citet{Augustine_2019}'s DDT study, while the initial conditions, refinement criteria, and explosion mechanism differ. These differences might not allow for a detailed realization-by-realization comparison, however a general comparison between the two explosion mechanisms can certainly be made.

\begin{figure}[ht!]
    \includegraphics[width=\columnwidth]{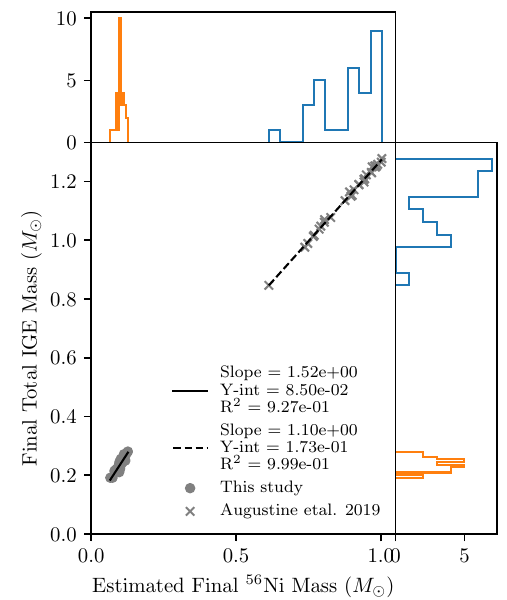}
    \caption{The final total \Ni{56} yield vs the final total IGE yield of this study with a deflagration explosion mechanism (grey $\bullet$) compared to \citet{Augustine_2019}'s study with a DDT explosion mechanism (grey $\times$).}
    \label{fig:augustine}
\end{figure}  

At the beginning of both simulations, the burning rate and IGE production closely follow each other, as the progenitor is undergoing a deflagration. The deflagration transitions into a detonation at about 1-2s of simulation time; realizations that burn less mass during the deflagration phase tend to transition sooner.

As a detonation incinerates the entire star, it is expected that our WD progenitor undergoing a DDT will have higher burned yields and explosion energies than when undergoing a deflagration alone. This large difference is shown in Figure \ref{fig:augustine}, which shows the disjoint ranges of final total IGE and \Ni{56} mass produced by the two explosion mechanisms. The DDT simulation, besides producing over 4 times greater yields, also has a much larger range of yields than its deflagration counterpart. Additionally, the \Ni{56}:IGE ratio of the DDT simulation is around 0.75, nearly twice than that of our deflagration simulation, 0.4.

\newpage
\subsection{Comparison to Observations} \label{subsec:obs}
Using Arnett's \edit1{model} \citep{Arnett_1982} or \citet{Valenti_2009}'s variation of it for a inner high-density region surrounded by a lower-density region, ejected masses can be calculated using parameters inferred from observations, including the ejecta velocity, opacity, and maximum brightness of the event. Table \ref{tab:obs} collects the results of these calculations performed by other studies for the dimmest Type Iax events. Where possible, data from multiple studies of the same Type Iax event is included. It is important to note that while there are over 70 members of the Type Iax subclass, there are only 13 events listed in Table \ref{tab:obs}. This is because generating lightcurves and spectra are enough information to classify an event as Type Iax, and analysis to determine the ejected masses and energies requires several additional assumptions and calculations. This search through the literature is not necessarily complete, but serves to demonstrates the range of brightnesses of Type Iax events: 0.001 - 0.30 \Msun ~ejected \Ni{56}.

\vspace{-30pt}
\begin{deluxetable*}{ccccccc}[ht!] \label{tab:obs}
\tablecaption{Ejecta yields calculated from observational data and other simulation studies. Data is ordered by increasing \Ni{56} mass while grouping studies of the same event.}
\tablehead{
\colhead{Reference} & \colhead{Type Iax} & \colhead{Magnitude} & \colhead{$v_{\rm{ej}}$} & \colhead{Mass} & \colhead{\Ni{56}} & \colhead{KE}\\
\colhead{} & \colhead{} & \colhead{} & \colhead{km/s} & \colhead{\Msun} & \colhead{\Msun} & \colhead{$10^{49}$ ergs} }
\startdata
\citet{SN2019gsc_Srivastav} & SN 2019gsc & ($M_r$) -13.97 $\pm$ 0.16 & 3500 & 0.13 - 0.22 & 0.0014 - 0.0024 & 1 - 2\\
\citet{2019gsc_Tomasella} & SN 2019gsc & ($M_r$) -14.28 $\pm$ 0.15 & 3500 & 0.13 & 0.003 $\pm$ 0.001 & 1.2\\
\citet{SN2008ha_Foley} & SN 2008ha & ($M_V$) -14.21 $\pm$ 0.15 & 2000 & 0.15 & 0.003 $\pm$ 0.0009 & 0.23\\
\citet{SN2008ha_Valenti} & SN 2008ha & ($M_R$) -14.5 $\pm$ 0.3 & 2300 & 0.1 - 0.5 & 0.003 - 0.005 & 1 - 5\\
\citet{SN2019gsc_Srivastav} & SN 2008ha &   &   & 0.1 - 0.2 & 0.003 & 1\\
\citet{2010ae_Stritzinger} & SN 2008ha & ($M_B$) -13.79 $\pm$ 0.14 & 4500 - 5500 & 0.4 - 0.5 & 0.004 & 8 - 18\\
\citet{SN2019gsc_Srivastav} & SN 2010ae &   & 3500$^a$ & 0.2 - 0.3 & 0.003 - 0.004 & 3 - 5\\
\citet{2010ae_Stritzinger} & SN 2010ae & ($M_B$) -13.79 $\pm$ 0.14 & 5000 - 6000 & 0.30 - 0.60 & 0.003 - 0.007 & 4 - 30\\
\citet{SN2020kyg_Srivastav} & SN 2020kyg & ($M_r$) -14.91 $\pm$ 0.15 & 3500$^a$ & 0.36$^{+0.08}_{-0.06}$ & 0.007 $\pm$ 0.001 & 4.4$^{+2.2}_{-1.5}$\\
\citet{SN2021fcg_Karambelkar} & SN 2021fcg & ($M_r$) -12.66 $\pm$ 0.20 &  & 0.05 - 0.4 & 0.008$^{+0.004}_{-0.005}$ &  \\
\citet{SN2007qd_McClelland} & SN 2007qd & ($M_B$) -15.4 $\pm$ 0.2 & 2800 & 0.15 & 0.013 $\pm$ 0.002 & 2.0\\
\citet{SN2019muj_Barna} & SN 2019muj & ($M_V$) -16.4 & 5000$^a$ & 0.17 & 0.031 $\pm$ 0.005 & 4.4\\
\citet{SN2015H_Magee} & SN 2015H & ($M_r$) -17.27 $\pm$ 0.07 & 5000-6000 & 0.5 & 0.06 &  \\
\citet{SN2014ck_Tomasella} & SN 2014ck & ($M_B$) -17.37 $\pm$ 0.15 & 3000 & 0.2-0.5 & 0.10$^{+0.04}_{-0.03}$ &  \\
\citet{SN2020rea_Singh} & SN 2020rea &  & 6500$^a$ & 0.77$^{+0.11}_{-0.21}$ & 0.13$\pm$ 0.01 & 19$^{+2}_{-6}$\\
\citet{SN2012Z_Stritzinger} & SN 2005hk & ($M_B$) -18.00 $\pm$ 0.25 & 5000 - 7000 & 1.5-2.0 & 0.15-0.20 & 40 - 120\\
\citet{SN2011ay_Szalai} & SN 2011ay & ($M_B$) -18.15 $\pm$ 0.17 & 6000 & 0.8 & 0.225 $\pm$ 0.010 & 10 \\
\citet{SN2012Z_Stritzinger} & SN 2012Z & ($M_B$) -18.27 $\pm$ 0.09 & 7000 - 9000 & 1.4-2.6 & 0.25-0.29 & 70 - 280\\
\hline
\citet{Kromer_2015}$^b$ & C/O/Ne WD & ($M_B$) -13.2 - -14.6 &   & 0.014 & 0.00340 & 0.18\\
\citet{Lach_2021}$^b$ & C/O WD & ($M_R$) -15.2 - -17.94 &   & 0.014 - 0.301 & 0.0058 - 0.092 & 0.18 - 9.7\\
\citet{Leung_2020}$^b$ & C/O/Ne WD &  & 5000 - 8000 & 0.96 - 1.30 & 0.18 - 0.36 &\\
\citet{Leung_2020}$^b$ & C/O WD &  & 5000 - 8000 & 0.92 - 1.36 & 0.23 - 0.34 &
\enddata
\tablecomments{Empty values indicate this quantity was not calculated.\\ $^a$Ejecta velocity was assumed as this value in order to perform calculations using Arnett's \edit1{model \citep{Arnett_1982, Valenti_2009}}.\\ ${^b}$These studies perform simulations of deflagrations. For more detail, see Section \ref{subsec:prevStudies}.  }
\end{deluxetable*}

In our simulation, it is assumed that all the bound material will eventually fall back and become part of the remnant, while the ejecta will move off into space to be detected by our instruments on Earth. Therefore, only the ejecta is considered when comparing to the yields inferred from observations. This is a reasonable assumption, because Arnett's \edit1{model} is only applied to data from the early light curve at maximum light, and it has been demonstrated that surface emission from the remnant only affects the late time evolution of the light curve \citep{Shen_2017}.

\begin{figure}[ht!]
    \includegraphics[width=\columnwidth]{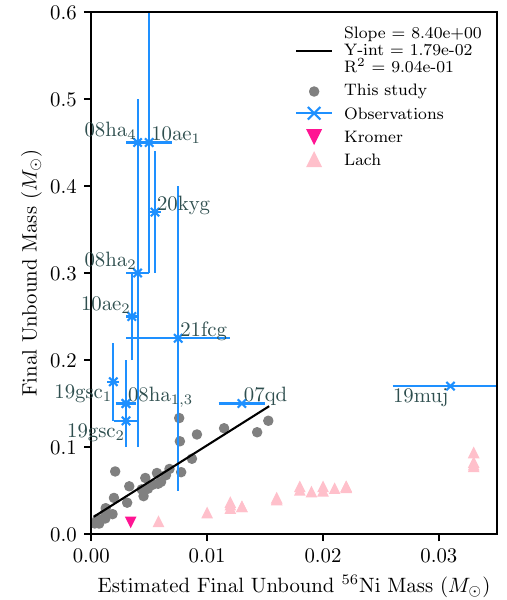}
    \caption{The final ejected mass vs the final ejected \Ni{56} mass of this study (grey $\bullet$) compared to observations (blue $\times$) and other studies: \citet{Kromer_2015} (fuscia $\triangledown$) and \citet{Lach_2021} (pink $\triangle$).}
    \label{fig:obs_ni}
\end{figure}

As shown in Figure \ref{fig:obs_ni}, our simulation produces \Ni{56} yields consistent with only the dimmest of the Type Iax population -- 
 SN 2019gsc \citep{SN2019gsc_Srivastav}, SN 2008ha \citep{SN2008ha_Foley}, SN 2010ae \citep{SN2010ae_Narayan}, SN 2020kyg \citep{SN2020kyg_Srivastav}, SN 2021fcg \citep{SN2021fcg_Karambelkar}, and SN 2007qd \citep{SN2007qd_McClelland}. These events have ejected \Ni{56} masses ranging from 0.0014 - 0.013 \Msun, while our simulation's range reahes an order of magnitude lower from 0.00030 - 0.01528 \Msun. \citet{Lach_2021}'s C/O WD deflagration study is unable to reach the yields of the very dimmest events, while \citet{Kromer_2015}'s C/O/Ne deflagration study comes close (0.0034 \Msun~ ejected \Ni{56}).

However, as with other studies of deflagrations, the ejected masses produced by our simulation (0.012 - 0.133 \Msun) are generally about 10 times lower that that inferred from the dimmest observational data. With the exception of the unusually large range calculated for SN 2021fcg, ejected masses are typically  $\geq$ 0.1 \Msun. This relationship is shown in Figure \ref{fig:obs_ni}.

\begin{figure}[ht!]
    \includegraphics[width=\columnwidth]{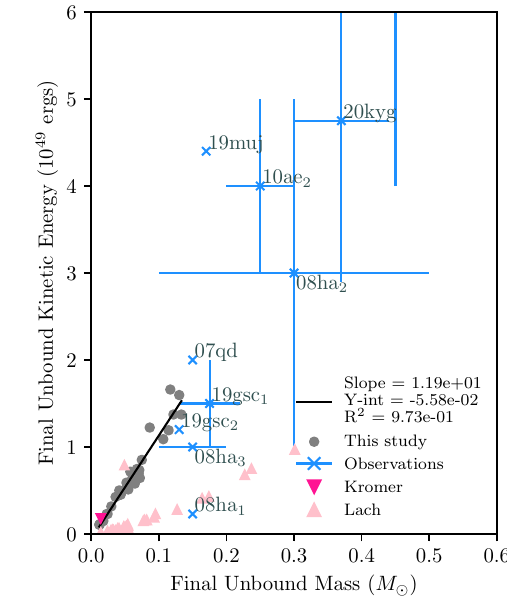}
    \caption{The final ejected KE vs the final ejected mass of this study (grey $\bullet$) compared to observations (blue $\times$) and other studies: \citet{Kromer_2015} (fuscia $\triangledown$) and \citet{Lach_2021} (pink $\triangle$).}
    \label{fig:obs_ke}
\end{figure}

When comparing the ejected KE and ejected masses of our simulation to observations (Figure \ref{fig:obs_ke}), it appears that our simulation reproduces the correct ratio, but at too low values. The question is then, is there a way to keep the \Ni{56} ejected mass the same, while increasing the total ejected mass and ejected KE of the system in the same ratio? This will be explored in a future 3D study, as discussed further in Section \ref{sec:conclusion}.

\newpage
\subsection{Core Reignition} \label{subsec:reignition}
It was noticed that in some realizations, during contraction of the central mass and fallback of material onto it, the remnant reached a high enough temperature to reignite. This is why we decided to manually turn off burning at 5s – the burning model described in Section \ref{subsec:flame} is not proper for modeling a core reignition. However, we decided to run a few of these realizations with only thermal burning ($\phi_{\rm{CC}}$) turned on after 3s. There are realizations that detonate, and ones that don’t. This brings the question as to how many realizations detonate, and what drives this decision. This is a difficult question to answer generally, and we do not attempt to answer it here, especially given the limits of our 2D simulations. However, this will be very much explored in our future 3D simulation suite, discussed in Section \ref{sec:future}.

\section{Discussion and Conclusion} \label{sec:conclusion}
This study performs \edit1{investigative} 2D hydrodynamic simulations of a deflagration in a \edit1{mixed} hybrid C/O/Ne WD to test its viability for explaining the dimmest of the Type Iax events. The simulation is performed in FLASH using a custom \edit1{3-stage burning} model based on the ADR scheme of \citet{Khokhlov_1995} that tracks the burning and energy release. The initial \edit1{hybrid} progenitor is the first created by a stellar evolution code (MESA) rather than a parameterized model, which is evolved from its initial ZAMS state through its accretion phase from a He-donor companion. To obtain a sample, we run 30 ``realizations" of the same simulation; the only difference between each realization is the number, shape, and amplitude of perturbations on the flame surface that is used to centrally ignite the star. \edit1{These realizations were chosen to replicate the initial flames in \citet{Augustine_2019}'s study of the same progenitor undergoing a DDT.}

Similar to other studies of deflagrations \edit1{\citep{Kromer_2015, Leung_2020, Lach_2021}}, our simulation's end state shows a dense bound remnant surrounded by sparse ejecta, where the majority of the system's mass is unburned. We obtain ejecta yields of: 0.012 - 0.133 \Msun ~total mass, 0.00117 - 0.04274 \Msun ~IGE,  0.00030 - 0.01528 \Msun ~\Ni{56}, and 0.107 - 1.660 $\times$ 10$^{49}$ ergs of KE. The remaining 1.23 - 1.35 \Msun ~is either in the dense core left behind, or will fall back onto this core to form a bound remnant. More detail of the yields of our simulation can be found in Table \ref{tab:yields}. We find that our simulation exhibits a characteristic ejecta velocity of 3500 km/s, which falls within the range of ejecta velocities inferred from observations \citep{Jha_2017}. We include only the unbound ejecta when comparing to observational data, however ongoing investigation seeks to determine if and how the trapped elements in the remnant will contribute to luminosity and spectroscopy measurements \citep{Shen_2017}.

Our study’s ejected \Ni{56} yields cover the range of the \Ni{56} yields of the dimmest of the Type Iax population. However, our total ejecta yields, which are about 10 times larger than the \Ni{56} yields, are smaller than the estimations provided by observational studies. This mismatch could stem from both the application of Arnett's \edit1{analysis} to observations, and the setup used in our simulation.

It is important to note that when estimating the \Ni{56} yields using Arnett’s \edit1{model}, there are assumptions made about a constant opacity and a central \Ni{56} distribution \citep{Arnett_1982, Valenti_2009}. As Arnett's \edit1{analysis} was created for use of explosions that fully disrupt the star, these assumptions may be less relevant for Type Iax events. However, Arnett's \edit1{model} still provides a valuable comparison to observations, and it is unlikely that it would overestimate ejecta yields by a full order of magnitude, while also having accurate \Ni{56} ejecta yields. Therefore, it will be important to generate measures such as light curves and spectra (see Section \ref{sec:future}) to make a better comparison with observations.

Differences in the progenitor (for example, metallicity), as well as different initial conditions of the model flame (for example, radius) may be able to produce a higher ejecta and KE yield in simulation, while keeping the same low \Ni{56} mass that our simulation produces. 
\edit1{Changing the initial matchhead radius and geometry (see Section \ref{subsec:initial}) was unable to achieve this desired outcome, however other initial conditions of the flame, such as placement, will be varied in a future study.} However, other studies of 3D deflagrations have explored this parameter space, the results of which will be summarized here. \citet{Lach_2021} found that increasing the central density of their parameterized C/O progenitor resulted in an increase in both IGE and KE yields, while the \Ni{56} yield increased to a point, but then decreased with increasing central density. This \edit1{decrease} is because electron capture rates are higher at larger densities, which results in a higher production of more stable isotopes of Ni. \citet{Lach_2021} also found that changing the metallicity of the initial ZAMS star did not have a systematic effect on the resulting ejecta composition or energy. Based on these results, it may be the case that a hybrid progenitor with a higher central density would have a higher total IGE and KE yield while keeping the \Ni{56} yield the same.

\edit1{As our progenitor is not a parameterized model, but rather a result of a full stellar evolution calculation, adjusting its central density and metallicity would require re-doing the calculations described in Section \ref{subsec:WD} with different assumptions.} Physically, the central density of a progenitor is a result of its isolation time - the time it takes for the star to get close enough to its companion to commence accretion. The longer this takes, the more time the progenitor has to cool and even crystallize, which increases its central density \edit1{at the time of ignition \citep{Lesaffre_2006}}. For us to create a hybrid progenitor with  a higher central density, it would be necessary to increase the distance between the WD and its companion in the MESA simulation described in Section \ref{subsec:WD}. However, there is no guarantee that doing this will create the desired hybrid WD, as hybrid WD have a tighter parameter space than traditional C/O ones. This would have to be done with much care.

Deflagrations of C/O/Ne WDs could explain the lowest brightness Type Iax events, but it is highly unlikely that this is the only method that produces a Type Iax event; these hybrid WDs are unusual and have a much lower C-fraction in the core, meaning there is less fuel to burn and turn into \Ni{56}. It may be that in the single degenerate paradigm, a range of compositions and densities of the initial progenitor may be able to explain the wide variation of Type Iax events, however the double degenerate paradigm should not be overlooked. Notably, as previously mentioned \edit1{in Section \ref{subsec:deflagration}}, the only observed Type Iax event remnant,``Parker's star'', is thought to be created by a C/O + O/Ne WD-WD merger \citep{Gvaramadze_2019, Oskinova_2020, Ritter_2021}. It has been suggested that studying more of these remnants may provide clues to determine the progenitor system of the event \citep{Ferrand_2021}. 

It is also uncertain how and why some of our realizations transition from a deflagration to a detonation upon contraction, incinerating the star and potentially producing a normal SN Ia. As the only difference between each realization is the initial ignition, it is interesting that some realizations detonate, while others do not (as far as we can simulate, $\sim$ 20s). This shows that the initial flame configuration is exceedingly important in determining the end state of the simulation, and must be carefully studied. Two applicable theories of reignition in a deflagration scenario are the pulsationally assisted gravitationally confined detonation (PGCD) \citep{Jordan_2012}, and the pulsating reverse detonation \citep{Bravo_2009}. Both mechanisms rely on the pulsation of the bound remnant as it contracts and pulls the surrounding material, which contains burning ashes, back towards it. The key difference between the two is where the detonation ignites: a PGCD will ignite the detonation inside the remnant's dense core where the burning ashes mix with the fuel, while a PRD ignites an inward-moving detonation at the interface of the remnant's core and the fallback ashes.

\citet{Lach_2022_reignition} also tested their models in the gravitationally confined detonation (GCD) scenario, where a deflagration wraps around the star and collides, sparking a detonation. This mechanism produced a much higher ejected \Ni{56} content (0.257 - 1.057 \Msun), with lightcurves and spectra most similar to 91T-like events, rather than Type Ia or Iax. 

\edit1{If reignition does not occur, the remnant will eventually form a compact object that is enriched with IGEs and C-burning ashes. It has been suggested that these object could be metallic hypervelocity ‘rogue stars’ that wander outside of galaxies as discussed by \citet{Hawkins_2015}, under the conditions of an asymmetric initial deflagration that gives the progenitor the kinetic energy required to escape its binary system. In 2D, our symmetry does not allow for an asymmetric deflagration, but it is possible and has been seen in a 3D configuration \citep{Kromer_2015, Lach_2021}. A potential candidate rogue star that may have originated from this type of progenitor system is LP 40-365, a metallic rogue star with a velocity of roughly 850 km/s that has been studied for decades. \citet{Hermes_2021} determined that based on the WD composition, the star likely had some type of thermonuclear origin, however, this does not explain its velocity. In theory, an asymmetric deflagration could explain the origin of hypervelocity stars like LP 40-365.}

\section{Future Work} \label{sec:future}

This 2D study has helped us identify, explore, and mitigate the challenges of evolving a deflagration in a hybrid progenitor, as well as shown the system to be a viable hypothesis for the dimmest Type Iax events that warrants further investigation. The most critical expansion of this study is to perform 3D simulations and to produce light curves to better compare with observational data. At the same time, exploring the parameter space is necessary to quantify the uncertainty in our results.

The initial flame radius is a free parameter in simulation that corresponds to how much the flame has evolved before the simulation is initialized. It is necessary to include this artificially burned initial region, as it is impossible to directly simulate a realistically sized flame ($\sim$ 1 cm scale) in a full-star ($\sim 10^8$ cm) simulation. It is important to determine the \edit1{effect} that this initial flame size \edit1{and placement} has on our results. Additionally, it will be important to test whether or not our simulations reignite, creating a delayed detonation and a much brighter event than a Type Iax. The relationship between the initial flame size and perturbations and the likelihood of detonation must be investigated.

Preparing a 3D simulation will involve modeling the pre-existing turbulence in the convective core and running the simulation until the ejecta are in free expansion. The former will be practically accomplished by introducing a turbulence field as in \citet{Jackson_2014}. The latter requires addressing the issue of potentially unphysical results from our EOS in the difficult low-temperature, low-density regions, discussed in Section \ref{subsec:separation}. An important step to solving this is to investigate the low-temperature area immediately in front of the burned plumes, to determine whether they are physically motivated or just a simulation artifact.

To produce light curves, Lagrangian tracer particles will be added to the simulation and post-processed in MESA to determine the elemental abundances for the simulation's end state, following the prescription of \citet{Townsley_2019}. These abundances will then be separated into those for remnant and ejecta. The ejecta abundances will be mapped to a velocity grid and a radiative transfer calculation will be performed to produce light curves and spectra. This will provide a much better comparison to observations than Arnett's \edit1{model}, although radiative transfer software introduce their own layer of assumptions.

More work needs to be done both in simulation and observational astronomy to understand these dim events. Normally, only the ejecta is used to create light curves to compare to observations, however it is unclear if and how the material in the bound remnant contributes. Some studies indicate that the late-time evolution of the light curve is affected by the remnant \citep{Shen_2017}. It is also unclear if a bound remnant has been observed in the aftermath of a Type Iax event, and this goal is impeded by the presence of the surviving companion in the binary system. Fortunately, there are active observational campaigns directed towards this goal, such as that of \citet{Jha_2021}. \edit1{Additionally, hypervelocity stars such as LP40-365 have been hypothesized to have a Type Iax origin \citep{Hermes_2021}.} Therefore, it is important to obtain as much detail as possible about the remnant that appears in simulations. Due to the difference in their evolution timescales, it is difficult to simultaneously resolve both the remnant and ejecta in simulations. Oftentimes the ejecta are focused on, so that light curves can be created for more direct comparison to observations. While evolving the remnant into a compact object in simulation will take work, it is an effort worth pursuing. Therefore, the goal is to run our simulation until the remnant's pulsation is calmer, and then the remnant abundances will be used as input to a stellar evolution calculation in MESA, to determine the makeup of the compact object that is produced in this process. This will allow a comparison to observations from the literature, and perhaps provide information for future observational campaigns.

\vspace{10pt}
\edit1{The authors thank the anonymous referee, whose input greatly enhanced the organization of this work. The authors also thank Ken Shen for many rewarding conversations and suggestions that improved this work.}

This work was supported in part by the US Department of Energy under grant DE-FG02-87ER40317 and by the National Science Foundation under grant 1927880. \edit1{Support also came from the Data + Computing =
Discovery summer REU program at the Institute for Advanced Computational Science at Stony Brook
University, supported by the NSF under award 1950052. This work
was also supported in part by the Simons Summer Research Program at Stony Brook University.}

The authors would like to thank Stony Brook Research Computing and Cyberinfrastructure, and the Institute for Advanced Computational Science at Stony Brook University for access to the high-performance SeaWulf computing system, which was made possible by a \$1.4M National Science Foundation grant (\#1531492). 

The software used in this work was developed in part by the DOE NNSA ASC- and DOE Office of Science ASCR-supported Flash Center for Computational Science at the University of Chicago. 

This research has made use of NASA’s Astrophysics Data System Bibliographic Services.

%


          
\software{MESA version 8118 (\url{https://mesa.sourceforge.net}) \citep{Paxton_2011, Paxton_2013, Paxton_2015}, 
          FLASH version 4.6.2 (\url{https://flash.rochester.edu}) \citep{Fryxell_2000, Calder_2002} with modifications for C/O and C/O/Ne WDs (\url{http://astronomy.ua.edu/townsley/code}), MESA-to-FLASH hydrostatic profile builder (\url{https://github.com/Flash-Star/mesautils}), \citep{Willcox_2016}, yt (\url{https://yt-project.org}) \citep{Turk_2010}
          }


\clearpage
\appendix
\section{Overview of the Burning Model and Equation of State} \label{ap:sim}
\subsection{Burning Model} \label{ap:flame}
In lieu of a full reaction network, the burning is separated into four groups of baryons ($\mathcal{X}_i$) that evolve in \edit1{three} stages ($\phi_{ij}$) as well as a dynamic \edit1{final NSE} state, the choice, energetics, and timescales of which are determined by both detailed ``self-heating'' models -- nuclear burning calculations performed at either constant pressure (isobaric) or constant density (isochoric) -- and calculations of steady-state detonations. These self-heating calculations used a 200 nuclide reaction network, and included the effects of electron screening and weak interactions \citep{Calder_2007}. The four groups of baryons are separated by their characteristic reactions: C and O fuel ($\mathcal{X}_f$); C-fusion ash ($\mathcal{X}_a$); Si-group element dominated nuclear statistical quasi-equilibrium (NSQE) ($\mathcal{X}_q$); and Fe-group element (IGE) dominated nuclear statistical equilibrium (NSE) ($\mathcal{X}_N$). The transitions between these four groups are described by corresponding $\phi$ variables: $\phi_{fa}$ tracks C-burning; $\phi_{aq}$ the conversion of C-burning ash to NSQE; and $\phi_{qn}$ the relaxation of NSQE to NSE. The properties of the final material are then evolved to account for changes due to adjustment of the NSE and electron capture. Here, $\mathcal{X}_i$ describe the baryon fraction of group $i$, and $\phi_{ij}$ describes the fraction of baryons that have undergone transition from state $i$ to the next consecutive state $j$ ($0 \leq \phi_{ij} \leq 1$). These are related by Equation 5 of \citet{Townsley_2016}:
\begin{equation} \label{eq:phi_vars}
    \phi_{fa} = \mathcal{X}_a + \mathcal{X}_q + \mathcal{X}_N,  \phi_{aq} = \mathcal{X}_q + \mathcal{X}_N,
    \phi_{qn} = \mathcal{X}_N
\end{equation}
with $1 \geq \phi_{fa} \geq \phi_{aq} \geq \phi_{qn}$.

These baryon fractions, rather than mass fractions, are evolved both in the burning model and in the hydrodynamics solvers, since baryon number and mass-energy are conserved in fusion reactions, rather than mass. Therefore, when discussing density, it is really the baryon density that is being referred to:
\begin{equation} \label{eq:baryondens} 
    \rho \equiv m_u n_B
\end{equation}
where $m_u$ is the atomic mass unit and $n_B$ is the baryon number density. However, the baryon density and mass density are similar enough in value to be used interchangeably when discussing the results of simulations \citep{Townsley_2009}.

\edit1{\citet{Willcox_2016} performed steady-state detonation calculations to determine the applicability of this three-stage burning model to an initial composition that includes \Ne{20}, a necessary nuclide in our hybrid progenitor. This was accomplished by running 225-nuclide steady-state Zel’doivch, von Neumann, and During (ZND) detonation models with varying mass fractions of \Ne{20} (0.01 - 0.45) -- the rest of the composition was chosen to be 0.02 \Ne{22}, 0.48 \Ox{16}, and the remainder \C{12}. \Ne{20} was consumed in the first stage of the burning, $\phi_{fa}$, with the sole effect being slowing the burning, while the rest of the stages were unaffected. Therefore, the only change required was to add \Ne{20} as an element in the first burning stage, to accurately represent changes in binding energy release and burning temperature. The three burning stages plus dynamic ash are described below.}

The burning model implemented in FLASH describes the consumption of C-fuel by considering contributions from both deflagration and thermally-activated detonation. The evolution of $\phi_{fa}$ is described by Equation 25 of \citet{Townsley_2016}:
\begin{equation}
    \frac{D\phi_{fa}}{Dt} = {\rm{max}}(0, r_{\rm{RD}}) + r_{\rm{CC}}
\end{equation}

$r_{\rm{RD}}$ is the contribution from the deflagration front, which is evolved by an advection-diffusion-reaction (ADR) equation which in general form may be written as
\begin{equation}
    \frac{D \phi}{Dt} = \frac{\partial \phi}{\partial t} +  \nabla \cdot (\vec{v}\phi) = \nabla \cdot(\kappa \nabla \phi) + \frac{1}{\tau}R(\phi)
\end{equation}
which describes the advection ($\nabla \cdot (\vec{v}\phi)$), diffusion ($\nabla \cdot(\kappa \nabla \phi)$), and reactions ($R(\phi)$) occuring on timescale $\tau$, of a substance $\phi$ in a fluid moving with velocity $\vec{v}$ and diffusivity $\kappa$.

The ADR equation for $\phi_{\rm{RD}}$, which represents the model flame, is presented in \citet{Townsley_2009}:
\begin{equation}
    r_{\rm{RD}} = \frac{D \phi_{\rm{RD}}}{Dt} = \kappa \nabla^2 \phi_{\rm{RD}} + \frac{1}{\tau}R(\phi_{\rm{RD}})
\end{equation}
In this form, the choices for the diffusion ($\kappa$) and reaction timescale coefficient ($\tau$) (described below) are specified to propagate a laminar flame at a given speed.
$ R(\phi)$ is described by a ``sharpened'' Kolmogorov Petrovski Piskunov (sKPP) equation \citep{Townsley_2007}:
\begin{equation}
    R(\phi) = \frac{1}{4} f (\phi - \epsilon_0)(1 - \phi + \epsilon_1)
\end{equation}
with $\epsilon_0$ = $\epsilon_1$ = $10^{-3}$ and $f$ = 1.309, constants that have been chosen to reduce noise in the solution and give the desired flame width and propagation speed in combination with other parameters defined here.

Following \citet{Townsley_2007}, the diffusion coefficient $\kappa$ is defined as:
\begin{equation}
    \kappa = \frac{1}{16} s b \Delta x
\end{equation}

and the reaction timescale coefficient $\tau$ as:
\begin{equation}
    \tau = \frac{1}{16} \frac{b \Delta x}{s} 
\end{equation}
where $b$ = 3.2, $s$ is the flame speed, and $\Delta x$ is the smallest cell size of the grid.

$r_{\rm{CC}}$ accounts for the contribution of thermal carbon-carbon
burning (ie in detonation), and is not stored as a variable. It is calculated from Equation 14 of \citet{Townsley_2009}: 
\begin{equation}
    r_{\rm{CC}} = \rho X_{^{12}C,f} (1-\phi_{fa})^2 \frac{1}{12} N_A \langle \sigma \nu \rangle_{C+C}(T) 
\end{equation}
where $\rho$ is the local baryon density defined in Equation \ref{eq:baryondens}, $X_{^{12}C,f}$ is the \C{12} fraction in pure fuel, and $N_A \langle \sigma \nu \rangle_{C+C}(T)$ is the temperature-dependent C-C fusion rate.

$r_{\rm{CC}}$ is introduced in the evolution of $\phi_{fa}$ to allow for a detonation to consume partially burned material at the artificially thick flame front. It is therefore suppressed ($r_{\rm{CC}} = 0$) where $\phi_{\rm{RD}} > 10^{-6} \land \phi_{fa} - \phi_{\rm{RD}} < 0.1$, to prevent thermally-activating burning in regions where the artificially thick flame is active, since the cell temperature in such regions is no longer representative of the actual burning process. As this work focuses on deflagration, $r_{\rm{CC}}$ is suppressed most of the time.

The other two $\phi$ variables are evolved at tuned timescales \citep{Townsley_2016}.
\begin{equation}
    \frac{D\phi_{aq}}{Dt} = \frac{\phi_{fa} - \phi_{aq}}{\tau_{\rm{NSQE}}(T)}
\end{equation}

\begin{equation}
    \frac{D\phi_{qn}}{Dt} = \frac{(\phi_{aq} - \phi_{qn})^2}{\tau_{\rm{NSE}}(T)}
\end{equation}

The inputs to these calculations, i.e. \edit1{the temperature-dependent timescales $\tau$ and the flame speed $s$} were determined as follows.

$\tau_{\rm{NSQE}}$ was defined by \citet{Calder_2007} using the aforementioned ``self-heating'' calculations, as the time taken for the system evolve from the point where 90\% of the \C{12} was burned, to the point of maximum \Si{28} abundance:
\begin{equation}
    \tau_{\rm{NSQE}}(T) = {\rm{exp}} (\frac{182.06}{T_9} - 46.054)
\end{equation}

$\tau_{\rm{NSE}}$ was tuned to the Si-group consumption timescale of a 200-nuclide, plane-parallel, steady-state Zel'doivch, von Neumann and During (ZND) detonation model \citep[][and references therein]{Townsley_2016}. This was accomplished by fitting the curves of Si-group consumption over time evaluated at pre-shock densities between 0.3 and 10 $\times$ 10$^7$ g/cm$^3$.
\begin{equation}
    \tau_{\rm{NSE}}(T) = {\rm{exp}} (\frac{201.0}{T_9} - 46.77)
\end{equation}

The flame speed, $s$, is dependent on the local composition and density. It is taken to be the maximum of the laminar flame speed, $s_{\rm{lam}}$, and a minimum flame speed, $s_{\rm{min}}$ that accounts for an increase in flame speed from the increased surface area of the flame as it undergoes the RT instability.
\begin{equation} \label{eq:flamSpeed}
    s = \rm{max}(s_{\rm{min}}, s_{\rm{lam}})
\end{equation}
where
\begin{equation} \label{eq:RTSpeed}
    s_{\rm{min}} = 0.5 \sqrt{Agm\Delta x}
\end{equation}
where $A$ is the C-burning Atwood number (between the unburned fuel and C-burning ash), $g$ is the acceleration due to gravity, $m$ is a tuneable parameter, $0.04 \leq m \leq 0.06$, and $\Delta x$ is the grid resolution.

$A$ is linearly interpolated from a table generated at the beginning of the simulation for a range of densities and \C{12}-fractions, using the global \Ne{22}-fraction. $s_{\rm{lam}}$ is similarly interpolated from a table of density, \C{12}-fraction, and \Ne{22}-fraction, which was obtained from detailed reaction network calculations by \citet{Chamulak_2007}, augmented by an interpolation of those performed by \citet{TimmesWoosley_1992} at lower densities \citep{Townsley_2009}. 
\edit1{For the hybrid model that includes \Ne{20}, $s_{\rm{lam}}$ is simply extrapolated to lower densities and carbon fractions. Since the flame propagation is dominated by the RT instability and turbulence, this is assumed to be a reasonable approximation \citep{Willcox_2016}.} 

Note for clarity that the above prescription for the flame speed is for 2D simulations \edit1{and does not include corrections for turbulence}. Future 3D simulations will use a background turbulent velocity field and appropriate corrections to speed from the turbulence-flame-interaction \citep{Jackson_2014}.

As seen in the direct numerical simulations of \citet{Calder_2007}, the realistic flame is on the order of $10^{-3}$ cm at the densities of the interior of the star ($10^9$g/cm$^3$). It is not computationally possible to reach this resolution while simultaneously resolving the WD, so the above description of the flame is practically implemented by artificially thickening it, so that the full range of $\phi_{fa}$ ([0,1]) always stretches across 4 grid zones. \citet{Townsley_2009} showed that the minimum resolution necessary to accurately describe the model flame was $\Delta x = $4 km. In reality, as the density of material burned by a flame decreases, the flame front widens, and this is described in our model by the energy release timescales $\tau_{\rm{NSQE}}$ and $\tau_{\rm{NSE}}$.

The final piece of the burning model is the adjustment of the properties of the material in NSQE and NSE, which is referred to as the dynamic ash state. While corresponding fusion and photo-dissociation (strong reactions) are held in balance, changes in density and temperature from the bulk fluid motion as well as neutronization from electron-capture (weak interactions) affect the equilibrium of this state. This dynamic ash state must therefore capture changes that occur on both the same timescales as the hydrodynamics and on longer timescales where weak interactions become important. To achieve this, the simulation evolves additional variables that track the electron fraction $Y_e$, the average number of nucleons per nucleus $\bar A$, and the average nuclear binding energy per baryon $\bar q$, which evolve both during the burning and in the final NSE state where they become especially important in influencing the amount of \Ni{56} produced \citep{Calder_2007}.

To determine how these variables change during the dynamic ash state, \citet{Seitenzahl_2009} expanded upon the work of \citet{Calder_2007} and performed calculations with 443 nuclides, which were then turned into tables. Tables for both isobaric and isochoric calculations are available for 0.44 $\leq Y_e \leq$ 0.50125, and using the set of values \{$Y_e$, $\rho$, $T$\}, $\dot Y_e$ and and the energy loss rate due to neutrinos ($\epsilon_\nu$) can be found by interpolation. From these, the resulting change in $Y_e$ and the rate of change in nuclear \edit1{binding} energy ($\epsilon_{\rm{nuc}}$) of the NSE state can be calculated, as described in \citet{Townsley_2007}:
\begin{equation}
    \dot Y_{e, \rm{NSE}} = \phi_{qn}\dot Y_e(Y_e, \rho, T)
\end{equation}

\begin{equation}
    \epsilon_{\rm{nuc}} = \frac{\Delta \bar q}{\Delta t} - \phi_{qn}[ \dot Y_e N_A c^2(m_p + m_e - m_n) + \epsilon_\nu ]
\end{equation}
where $N_A$ is Avogadro's number and $m_p$, $m_e$, and $m_n$ are the mass of the proton, electron, and neutron, respectively.

\subsection{Equation of State} \label{ap:eos}
The EOS used for our application is the Helmholtz EOS described in \citet{TimmesAndSwesty_2000} with additional Coulomb corrections. It calculates pressure ($P$), internal energy ($E_{\rm{int}}$), and entropy ($S$) contributions separately for radiation, ions, electrons, positrons, and Coulomb effect corrections, and sums them together. For radiation, a thermal photon distribution is used whose temperature matches that of the other components; the ions are assumed to act as an ideal gas; the electrons and positrons are modeled as a non-interacting Fermi-gas; and the Coulomb corrections account for the interaction between the ions and free electrons. Calculating the state variables of the Fermi-gas requires evaluating the chemical potential ($\eta$) using the Fermi-Dirac integrals. The material is assumed to be fully ionized, though no checks are made to avoid regions where that is known to be inaccurate, as doing so in regions where pressure ionization is present, rather than the usual Saha ionization, does not have a simple well-defined procedure. More broadly applicable equations of state typically use tabulations in such regions, e.g. the EOS used in the MESA \citep{Jermyn_2023}.

In the Helmholtz EOS, the Helmholtz free energy, $F$, and eight of its partial derivatives are tabulated as a function of density and temperature. $P$, $E_{\rm{int}}$, and $S$ are then computed from this lookup table. This table was computed using the Timmes EOS, which was designed to be accurate at the cost of being computationally intensive \citep{TimmesAndArnett_1999}. The Timmes EOS calculated $\eta$ to 18 significant figures using a Newton-Raphson method, evaluated the Fermi-Dirac integrals exactly in IEEE 64-bit arithmetic, and analytically solved for the derivatives of pressure and thermal energy. To keep this accuracy (IEEE 64-bit precision) but reduce the computation time (by over a factor of 100), these results were used to calculate $F$ and its partial derivatives, which were then tabulated \citep{TimmesAndSwesty_2000}. The granularity of the table affects the accuracy of the result -- we used the same table as that in CASTRO \citep{Almgren_2010}, which has values for 271 densities and 101 temperatures in the limits $10^{-10} < \rho < 10^{11}$ g/cm$^3$ and $10^4 < T < 10^{11}$ K\footnote{This extended table is available at: \url{https://github.com/AMReX-Astro/Microphysics/blob/main/EOS/helmholtz/helm_table.dat}}. Although this table was created with $Y_e = 1$ (pure hydrogen), it can be directly applied to any $Y_e$ by making the appropriate adaptations based on $Y_e$.

 The choice to tabulate $F$ and its derivatives is dictated by the desire for tight thermodynamic consistency and the use of temperature and density as the desired independent variables, therefore making Helmholtz potential the corresponding fundamental Legendre-transformed potential. $P$, $E_{\rm{int}}$, and $S$ can be easily constructed from $F$ and its derivatives: 
\begin{equation}
    P = \rho^2 \left( \frac{\partial F}{\partial \rho} \right)_T
\end{equation}

\begin{equation}
    E_{\rm{int}} = F + TS
\end{equation}

\begin{equation}
    S = -\left(\frac{\partial F}{\partial T} \right)_\rho
\end{equation}
where the value in the subscript is held constant for the thermodynamic partial
derivative.

A biquintic polynomial is used to interpolate the table values in 2D ($\rho$, $T$) space between grid cells, to ensure both that thermodynamic consistency is maintained and that the derivatives of $P$, $\epsilon_{\rm{int}}$, and $S$ are continuous across the grid.

\clearpage
\typeout{} 
\bibliography{bibliography}{}
\bibliographystyle{aasjournal}



\end{document}